\documentclass[%
 reprint,
nofootinbib,
amsmath,
amssymb,
aps,
prd,
floatfix,
twocolumn
]{revtex4-2}

\usepackage{graphicx}
\usepackage{dcolumn}
\usepackage{bm}
\usepackage{booktabs}
\usepackage{colortbl}
\usepackage{braket} 
\usepackage{amsmath}
\usepackage{float}
\usepackage{multirow}
\usepackage{hyperref}
\usepackage[dvipsnames]{xcolor}
\usepackage[compat=1.1.0]{tikz-feynman}
\usepackage[compat=1.1.0]{tikz-feynhand}
\hypersetup{
    colorlinks,
    linkcolor={Emerald},
    citecolor={Emerald},
    urlcolor={Emerald}
}
\usepackage{tikz}
\usepackage[compat=1.1.0]{tikz-feynman}
\usepackage[compat=1.1.0]{tikz-feynhand}
\usepackage[export]{adjustbox}
\usepackage{color, colortbl}

\newcommand{\sig}{\sigma}
\newcommand{\lzm}{\left(}
\newcommand{\dzm}{\right)}
\newcommand{\lzs}{\left[}
\newcommand{\dzs}{\right]}
\newcommand{\lzv}{\left\{}
\newcommand{\dzv}{\right\}}

\newcommand{\cL}{\mathcal{L}}
\newcommand{\cO}{\mathcal{O}}
\newcommand{\cR}{\mathcal{R}}

\newcommand{\cW}{{\mathcal W}}
\newcommand{\cB}{{\mathcal B}}

\newcommand{\cC}{{\mathcal C}}
\newcommand{\cS}{{\mathcal S}}

\newcommand{\hermc}{\text{h.c.}}

\newcommand{\eminus}{\vcenter{\hbox{\scalebox{0.6}[1]{$ - $}}}}	

\newcommand{\sscript}[1]{{\scriptscriptstyle \mathrm{#1}}}

\begin{document}
\title{Discrete Leptonic Flavor Symmetries: UV Mediators and Phenomenology}

\author{Ajdin Palavri\'{c}}
 \email{ajdin.palavric@unibas.ch}

\affiliation{%
 Department of Physics, University of Basel, Klingelbergstrasse 82,  CH-4056 Basel, Switzerland
}%

\begin{abstract}

Given the absence of a definitive top-down indication for understanding the peculiar structure of the lepton sector, discrete flavor symmetries offer a profound perspective for examining the intricate patterns of lepton masses and mixings. In this work, drawing upon previous studies on the interplay of flavor symmetries with the potential UV completions from a purely bottom-up perspective, three well-motivated discrete flavor groups, suitable for portraying the leptonic sector as well as the neutrino masses, specifically $A_4$, $A_5$ and $S_4$, are explored within this framework, leading to the comprehensive classification of the NP mediators, along with the tree-level matching relations onto dimension-6 SMEFT operators. Particular emphasis is placed on the \textit{discrete leptonic directions}, for which a phenomenological analysis is carried out in order to constrain various NP mediators, where significant focus is directed towards the examination of the cLFV operators, which, for the wide range of applicable cases, offer the leading constraint.


\end{abstract}

\maketitle

\section{\label{sec:setup}Introduction}

Amidst the lack of direct observation of new physics (NP) states, which points to a discernible scale separation between the electroweak (EW) scale and the scale NP mediators could be associated to, establishing a reliable and rigorous theoretical framework with the aim of systematically characterizing deviations from the Standard Model (SM) is a task of the utmost significance. Such theoretical framework emerges in the form of the Standard Model Effective Field Theory (SMEFT)~\cite{Buchmuller:1985jz, Grzadkowski:2010es, Brivio:2017vri, Isidori:2023pyp, Giudice:2007fh, Henning:2014wua}. The SMEFT Lagrangian can be expressed as an infinite series of local higher-dimensional operators whose canonical mass dimension is greater than 4. These local higher-dimensional operators are constructed using the SM fields and demanding the invariance under the gauge and Poincaré symmetry. All of the operators from the series are accompanied by the respective Wilson coefficients (WCs), which encompass the short-distance effects and are, a priori, unknown and don't rely on any particular model generating them, therefore providing a robust model-independent framework suitable for application in a variety of phenomenological studies.

The prevalent trait of the SMEFT Lagrangian is the fact that at each order in the inverse power of the cutoff scale there is a finite number of independent operators of given canonical mass dimension that can be constructed. This feature inspired and drove a significant advancement across many aspects of SMEFT including e.g. operator counting~\cite{Henning:2015alf, Fonseca:2019yya} and operator basis construction for various canonical dimensions~\cite{Grzadkowski:2010es, Lehman:2014jma, Li:2023cwy, Ren:2022tvi, Li:2020xlh, Li:2020gnx, Murphy:2020rsh, Liao:2016hru, Liao:2020jmn, Harlander:2023psl, Harlander:2023ozs}. However, in spite of the fact that each order of the expansion is characterized by the finite number of higher-dimensional operators, increasing the mass dimension leads to rapid proliferation in the number of independent operators. This proliferation, along with being caused by the increase in number of possible operator structures allowed by the gauge and Poincaré symmetries, is mainly ascribed to the fact that we have three generations (flavors) of the fermionic fields. To exemplify this point, it suffices to consider the baryon-number conserving operators at dimension six, where, in presence of a single active flavor, the number of independent parameters is 59, whereas, if all three flavors are active, the number of independent parameters grows to 2499~\cite{Alonso:2013hga}.

Nevertheless, at the level of the SM, in absence of the Yukawa interactions, the kinetic Lagrangian for the five different fermionic gauge representations is invariant under $U(3)^5$ flavor symmetry. This large global symmetry is broken upon introduction of the Yukawa interactions down to the global baryon and three individual lepton numbers. Despite the breaking of the $U(3)^5$ symmetry, the empirical evidence arising from the observed quark masses and mixing angles remains to hint at the existence of an approximate flavor symmetry, which could be used to portray the observational input. 

Additionally, from the point of view of BSM, stringent constraints are imposed on NP as a result of the precise flavor experiments~\cite{EuropeanStrategyforParticlePhysicsPreparatoryGroup:2019qin} and with the aim of having the NP within the reach of the present or future experiments (not far from the TeV range), the approximate flavor symmetries should not be excessively violated. 

Combining this phenomenological reasoning with the fact that the quick proliferation of the independent parameters in the SMEFT mainly emanates due to the existence of flavor, a natural next step entails imposing the approximate flavor symmetries at the level of SMEFT operators, which undeniably offers valuable insights into the structure of the SMEFT landscape~\cite{Greljo:2022cah, Faroughy:2020ina, Bruggisser:2022rhb, Machado:2022ozb}. 

Moreover, in line of the bottom-up approach, one pertinent topic worth exploring is related to identifying the NP mediators in a model-independent way, determining their possible interactions with the SM degrees of freedom and answering the question which mediators are capable of generating which SMEFT operators, once integrated out at tree level. The complete spectrum of these NP fields including scalar, fermion and vector mediators and the corresponding tree-level matching relations to dimension-6 SMEFT operators is presented in Ref.~\cite{deBlas:2017xtg}.

In consideration of the preceding remarks regarding flavor symmetries, the exercise of imposing the flavor symmetry can also be extended to the NP mediators themselves and the impact on the tree-level matching relations can then be examined accordingly. Analysis of this type, performed imposing an exact $U(3)^5$ flavor symmetry was conducted in Ref.~\cite{Greljo:2023adz}. In addition to the full classification of the NP mediators based on the possible flavor irreducible representations (irreps), a key feature of that analysis was that the matching conditions for a vast majority of NP flavor irreps were given as a well-defined linear combination of dimension-6 SMEFT operators multiplied by a single parameter. The matching conditions of that form were referred to as \textit{leading directions}. Given that the leading directions contained just a single parameter, the bounds on the scale for many of the NP flavor irreps could be estimated by means of a simple, tree-level phenomenological analysis. In addition, going beyond the tree level and studying the RG effects proves to be impactful for certain directions as well, as noted in Ref.~\cite{Greljo:2023bdy}.

Redirecting our attention to the flavor symmetry groups examined in Ref.~\cite{Greljo:2022cah}, one can observe that all of the considered ones were global continuous groups. Furthermore, unlike in the quark sector, where the pronounced hierarchy between the third and the first two generations already indicates several suitable choices for the flavor symmetries, such as $U(2)^3$, on the leptonic side, however, there are numerous additional viable options worth considering.

In this letter, focusing solely on the leptonic sector, we broaden the list of symmetry groups considered in Ref.~\cite{Greljo:2022cah} by exploring three discrete flavor symmetry groups, which we directly analyze in the context of the aforementioned directions~\cite{Ishimori:2010au}. Driven by the comprehensive body of work on examining the role of the discrete symmetries in relation to the lepton sector, we will center our analysis to three discrete groups: $A_4$~\cite{Morisi:2009sc, Ferreira:2013oga, Ding:2019zxk, Feruglio:2008ht, Holthausen:2012wz, Abbas:2020qzc, Morisi:2013eca, DeLaVega:2018bkp, Peinado:2010iu, Pramanick:2015qga, Altarelli:2009kr, Pramanick:2019qpg, Carrolo:2022oyg, Ding:2021eva, King:2011ab, Ding:2024fsf, Kadosh:2010rm, Barman:2023idm, Borah:2018nvu, Boucenna:2011tj, Memenga:2013vc, Morisi:2007ft, Nomura:2019jxj, Zhang:2019ngf, Asaka:2019vev, Kumar:2023moh, Okada:2019mjf, Gogoi:2022jwf, Singh:2023jke, Kobayashi:2018scp, Criado:2019tzk, Criado:2018thu, Feruglio:2017spp, Altarelli:2010gt, CentellesChulia:2023osj, Kumar:2024zfb, Chauhan:2023faf, Borah:2017dmk, Karmakar:2014dva, Feruglio:2019ybq, Petcov:2018snn, King:2013eh}, $A_5$~\cite{Ding:2011cm, Novichkov:2018nkm, DiIura:2016ols, DiIura:2015kfa, Cooper:2012bd, Ding:2019xna, Li:2015jxa, Turner:2015mwa, Puyam:2023div, Gehrlein:2014wda, Hernandez:2012ra, deMedeirosVarzielas:2022ihu, Ballett:2015wia, Feruglio:2011qq, Everett:2008et, Gehrlein:2015dza, Gehrlein:2015dxa, Chen:2010ty} and $S_4$~\cite{Kobayashi:2019mna, Ishimori:2009ns, Thapa:2023fxu, Ishimori:2011mt, Hagedorn:2006ug, Bazzocchi:2012st, Zhang:2021olk, Ishimori:2010fs, Izawa:2023qay, Bazzocchi:2008ej, Wang:2019ovr, Nomura:2021ewm, Penedo:2018nmg, Novichkov:2018ovf, King:2021fhl, Ding:2019gof, Okada:2019lzv}, which stand out as favorable candidates for explaining its flavor structure. Furthermore, as evidenced in the aforementioned references, these discrete flavor symmetries play a significant role in the exploration of various models addressing the neutrino mass textures and mixing angles. 

Before advancing further, two brief points should be noted: first, analogously to the analysis presented in Refs.~\cite{Kobayashi:2021pav, Kobayashi:2023zzc}, in this work the discrete flavor symmetries are considered below the supersymmetry (SUSY) breaking scale and second, along the same lines of reasoning as in Ref.~\cite{Greljo:2023adz}, where the exact $U(3)^5$ flavor symmetry is imposed at the level of the interaction Lagrangians of the NP mediators with the SM fields, in the forthcoming analysis the similar approach is employed taking the exact discrete symmetries.\footnote{In accordance with Refs.~\cite{Greljo:2023adz, Singh:2023jke}, imposing the exact discrete flavor symmetries corresponds to capturing the leading-order effects in the flavor power counting, i.e. in absence of flavor spurions. In order to account for the flavor power counting beyond the leading order, the discrete flavor symmetry groups can be promoted to modular symmetries, which introduces the Yukawa matrices as the modular forms, see e.g. Ref~\cite{Kobayashi:2021pav}.}

As described in Ref.~\cite{deBlas:2017xtg}, we can isolate 13 NP mediators, whose interaction Lagrangians contain terms which designate the coupling of the NP mediators to the lepton fields: 4 scalars ($\cS_1$, $\cS_2$, $\varphi$, $\Xi_1$), 6 fermions ($N$, $E$, $\Delta_1$, $\Delta_3$, $\Sigma$, $\Sigma_1$) and 3 vectors ($\cB$, $\cW$, $\cL_3$). The gauge irreps, written in the $(SU(3)_C,SU(2)_L)_{U(1)_Y}$ format, and the relevant terms of the interaction Lagrangians for these 13 mediators are listed in Tab.~\ref{tab: intro mediators and int}. Our principal objective is to perform the full classifications of these 13 NP mediators imposing $A_4$, $A_5$ and $S_4$ flavor symmetries and, throughout the process of finding the tree-level matching relations onto dimension-6 SMEFT operators, study the phenomenology of the \textit{discrete leptonic directions}.\footnote{Among the 13 NP mediators considered, certain ones, such as $N$, $\varphi$ and $\cS_{1,2}$, are also featured in various neutrino mass models in presence of discrete flavor symmetries (see e.g. Refs.~\cite{Ferreira:2013oga, Holthausen:2012wz, Abbas:2020qzc}).}

This letter is structured as follows: in Sec.~\ref{sec:discrete_flavor_symmetries} we give a brief overview of three discrete flavor symmetries considered, focusing in particular on the representation theory of each group. In an effort of extracting the flavor structure of the coupling tensors, we provide expressions for relevant tensor product expansions including relations written in the component notation for significant instances.\footnote{Coupling tensors in presence of flavor symmetries are referred to as \textit{flavor tensors} in the remainder of the present work.} After this point, we determine the possible flavor irreps for all NP mediators and extract the flavor tensors. In Sec.~\ref{sec:directions_and_pheno} we perform the tree-level matching onto dimension-6 SMEFT operators defining the discrete leptonic directions. Subsequent to this, we perform a phenomenological analysis, similar to one presented in Ref.~\cite{Greljo:2023adz}. The lower bounds on different irreps are found by considering the low-energy observables as well as the constraints from the cLFV SMEFT operators. Lastly, in Sec.~\ref{sec:conclusions} we present the summary and offer hints of promising directions for future study.

\begin{table}[t]
\centering
\begin{tabular}{cc}
\toprule
\textbf{UV Field} & \textbf{$-\cL_{\sscript{UV}}^{(4)}\supset$}
\\
\midrule
$\cS_1\sim(\bm1,\bm1)_{1}$
&$[y_{\mathcal S_1}]_{rij}\mathcal S_{1r}^\dag \bar\ell_{i}i\sig_2\ell_{j}^c$
\\
$\cS_2\sim(\bm1,\bm1)_{2}$
&$[y_{\mathcal S_2}]_{rij}\mathcal S_{2r}^\dag \bar e_{i}e_{j}^c$
\\
$\varphi\sim(\bm1,\bm2)_{\frac{1}{2}}$
&$[y_\varphi]_{rij}\varphi_r\bar\ell_ie_j$
\\
$\Xi_1\sim(\bm1,\bm3)_{1}$
&$[y_{\Xi_1}]_{rij}\Xi_{1r}^{a\dag}\bar\ell_{i}\sig^ai\sig_2\ell_{j}^c$
\\
\midrule
\midrule
$N\sim(\bm1,\bm1)_0$
&$[\lambda_N]_{ri}\bar N_{R,r}\tilde\phi^\dag \ell_i$
\\
$E\sim(\bm1,\bm1)_{-1}$
&$[\lambda_E]_{ri}\bar E_{R,r}\phi^\dag\ell_{i}$
\\
$\Delta_1\sim(\bm1,\bm2)_{-\frac{1}{2}}$
&$[\lambda_{\Delta_1}]_{ri}\bar\Delta_{1L,r}\phi e_i$
\\
$\Delta_3\sim(\bm1,\bm2)_{-\frac{3}{2}}$
&$[\lambda_{\Delta_3}]_{ri}\bar\Delta_{3L,r}\tilde\phi e_{i}$
\\
$\Sigma\sim(\bm1,\bm3)_{0}$
&$\frac{1}{2}[\lambda_\Sigma]_{ri}\bar\Sigma^a_{R,r}\tilde\phi^\dag\sig^a\ell_{i}$
\\
$\Sigma_1\sim(\bm1,\bm3)_{-1}$
&$\frac{1}{2}[\lambda_{\Sigma_1}]_{ri}\bar\Sigma^a_{1R,r}\phi^\dag\sig^a\ell_{i}$
\\
\midrule
\midrule
$\cB\sim(\bm1,\bm1)_0$
&$[g_\cB^\ell]_{rij}\cB_r^\mu\bar\ell_{i}\gamma_\mu\ell_{j}+[g_\cB^e]_{rij}\cB_r^\mu\bar e_{i}\gamma_\mu e_{j}$
\\
$\cW\sim(\bm1,\bm3)_0$
&$\frac{1}{2}[g_\cW^\ell]_{rij}\cW^{\mu\,a}_r\bar\ell_i\sigma^a\gamma_\mu\ell_j$
\\
$\cL_3\sim(\bm1,\bm2)_{-\frac{3}{2}}$
&$[g_{\cL_3}]_{rij}\cL_{3r}^{\mu\dag}\bar e_{i}^c\gamma_\mu \ell_{j}$
\\
\bottomrule
\end{tabular}
\caption{Overview of the NP mediators considered in this work. The gauge irreps are given in the first, whereas the relevant terms of the interaction Lagrangians are indicated in the second column. Hermitian conjugates are implicitly assumed where appropriate. The interaction terms are written in the most general way, i.e. in absence of any flavor symmetries. $\phi$ denotes the SM Higgs field and $\sigma^a$ Pauli matrices, where $a=1,2,3$. $\tilde\phi$ is defined in the standard way as $\tilde\phi=i\sigma^2\phi^*$.}\label{tab: intro mediators and int}
\end{table}

\section{Discrete flavor symmetries and flavor tensors}
\label{sec:discrete_flavor_symmetries}
In this section we focus on providing more technical details for the $A_4$, $A_5$ and $S_4$ flavor symmetry groups, paying particular attention to the representation theory and tensor product expansions. In the following step, we specify the flavor irreducible representations of the SM fields, which enable us to extract the flavor tensors for all UV mediators collected in Tab.~\ref{tab: intro mediators and int}. 


\subsection{$A_4$ flavor symmetry}

\subsubsection{Overview of the group properties}
$A_4$ group is a finite, alternating group on four objects, which consists of all even permutations. $A_4$ symmetry group contains $4!/2=12$ elements. Geometrically, $A_4$ is isomorphic to the symmetry group of a regular tetrahedron, which implies that this group describes the rotational symmetries of a regular tetrahedron, excluding reflections. 

The common attribute shared by the discrete flavor groups discussed in the present work is that these groups can be generated by two generators, commonly labeled as $S$ and $T$ generators, with the following property, see e.g. Ref.~\cite{Feruglio:2008ht}
\begin{equation}\label{eq: A4 definition}
    S^2=T^3=(ST)^3=1.
\end{equation}

In terms of the representation theory, $A_4$ proves to be the smallest finite non-Abelian group which allows for a triplet representation ($\bm3$), along with three distinct singlets, which are denoted as $\bm1$, $\bm1'$ and $\bm1''$. In order to construct the tensor products, explicit forms of the $S$ and $T$ generators need to be specified and this is especially relevant for the nontrivial irreps, such as $\bm3$ in the case of $A_4$.

For three $A_4$ singlets the $S$ and $T$ generators can be written as
\begin{equation}
    \begin{alignedat}{5}
        \bm{1}&: &\qquad S&=1,&\qquad T&=1,
        \\
        \bm{1}'&: &\qquad S&=1,&\qquad T&=\omega,
        \\
        \bm{1}''&: &\qquad S&=1,&\qquad T&=\omega^2,
    \end{alignedat}
\end{equation}
where $\omega=e^{2\pi i/3}=-1/2+i\sqrt{3}/2.$ It is evident that this arrangement satisfies the defining relation given by Eq.~\eqref{eq: A4 definition} along with the fact that the tensor product expansions for the singlets become
\begin{equation}
    \bm1\otimes\bm1=\bm1,\quad \bm1'\otimes\bm1'=\bm1'',\quad \bm1''\otimes\bm1''=\bm1',\quad \bm1'\otimes\bm1''=\bm1.
\end{equation}
In order to find the decomposition of the product involving two $A_4$ triplet representations, the $S$ and $T$ generators can be written as $3\times3$ matrices, where we can opt for the basis in which one of these generators is diagonal:
\begin{equation}
    S=\frac{1}{3}
    \begin{bmatrix}
        -1&2&2\\2&-1&2\\2&2&-1
    \end{bmatrix},
    \qquad
    T=\begin{bmatrix}
        1&0&0\\0&\omega&0\\0&0&\omega^2
    \end{bmatrix},
\end{equation}
as demonstrated e.g. in Refs.~\cite{Ishimori:2010au, Kobayashi:2021pav, Feruglio:2008ht, Ding:2019zxk}. Bearing this in mind, the tensor product involving two $A_4$ triplets becomes
\begin{equation}\label{eq:A4_tensor_product}
    \bm3\otimes\bm3=\bm1\oplus\bm1'\oplus\bm1''\oplus\bm{3_S}\oplus\bm{3_A},
\end{equation}
where $S$ and $A$ indicate the symmetric and antisymmetric combinations of the triplet components, respectively. Taking $\alpha\equiv(\alpha_1,\alpha_2,\alpha_3)^T\sim\bm3$ and $\beta\equiv(\beta_1,\beta_2,\beta_3)^T\sim\bm3$, the irreps obtained as a result of taking the tensor product involving two $A_4$ triplets can be expressed in terms of the components of $\alpha$ and $\beta$. The singlets in the decomposition given by Eq.~\eqref{eq:A4_tensor_product} are given as\footnote{We use the notation of the form $[\mathcal{R}_1\mathcal{R}_2]_{\mathcal{R}_3}\equiv
    (\mathcal{R}_1\otimes\mathcal{R}_2)_{\mathcal{R}_3}$, where $\mathcal{R}_1$ and $\mathcal{R}_2$ are the irreps entering the tensor product and $\mathcal{R}_3$ is a single irrep we pick from the tensor product expansion.}
\begin{equation}\label{eq:A4_singlets}
    \begin{alignedat}{2}
        [\alpha\beta]_{\bm1}&=\alpha_1\beta_1+\alpha_2\beta_3+\alpha_3\beta_2,
        \\
        [\alpha\beta]_{\bm1'}&=\alpha_3\beta_3+\alpha_1\beta_2+\alpha_2\beta_1,
        \\
        [\alpha\beta]_{\bm1''}&=\alpha_2\beta_2+\alpha_1\beta_3+\alpha_3\beta_1,
    \end{alignedat}
\end{equation}
while the symmetric and antisymmetric triplet combinations can be written as
\begin{equation}\label{eq:A4_triplets}
    \begin{alignedat}{2}
        [\alpha\beta]_{\bm{3_S}}&=\frac{1}{3}
        \begin{bmatrix}
            2\alpha_1\beta_1-\alpha_2\beta_3-\alpha_3\beta_2
            \\
            2\alpha_3\beta_3-\alpha_1\beta_2-\alpha_2\beta_1
            \\
            2\alpha_2\beta_2-\alpha_1\beta_3-\alpha_3\beta_1
        \end{bmatrix},
        \\
        [\alpha\beta]_{\bm{3_A}}&=\frac{1}{2}
        \begin{bmatrix}
            \alpha_2\beta_3-\alpha_3\beta_2
            \\
            \alpha_1\beta_2-\alpha_2\beta_1
            \\
            \alpha_3\beta_1-\alpha_1\beta_3
        \end{bmatrix}.
    \end{alignedat}
\end{equation}
Using these equations we can take the next step and extract the flavor tensors for the NP mediators from Tab.~\ref{tab: intro mediators and int}.

\subsubsection{Extraction of flavor tensors}
Having reviewed some of the basic properties of the $A_4$ group and upon specifying the irreducible representations and the tensor products, our next step entails determining the possible irreducible representations of the NP mediators and extracting the flavor tensors from their corresponding UV interaction terms.

Regarding the determination of the possible irreps for the UV mediators, we largely adhere to the strategy outlined in Ref.~\cite{Greljo:2023adz}, where the flavor irreps of the UV mediators are determined by specifying the flavor irreps for the SM fields and imposing the invariance of the UV Lagrangian under the chosen flavor symmetry.

Under this particular flavor symmetry, we take the $SU(2)_L$ lepton doublet to transform as an $A_4$ triplet, while three $A_4$ singlets are associated to three generations of the right-handed lepton singlets, which is the common choice in various models involving the (modular) $A_4$ symmetry, e.g. as demonstrated in Refs.~\cite{Kobayashi:2021pav, Feruglio:2008ht, Ding:2019zxk, Abbas:2020qzc, Pramanick:2015qga}:\footnote{Conjugated spinors follow the same arrangement, while $\bar\ell$ and $\bar e$, in case of $\cB$ and $\cW$, can be taken with the arrangement outlined in Ref.~\cite{Kobayashi:2021pav} with the second and third component of the triplet flipped in order to make the kinetic term diagonal in flavor.}
\begin{equation}
    \ell\equiv(\ell_1,\ell_2,\ell_3)^T\sim\bm3,\quad (e_1,e_2,e_3)\sim(\bm1,\bm1',\bm1'').
\end{equation}

With the flavor irreps of the SM fields specified, accompanied by the $A_4$ tensor products, we can determine which flavor irreps can be associated to the UV mediators such that the UV interaction Lagrangians are kept invariant. The complete $A_4$ classification of the UV mediators, along with their UV interaction terms, $A_4$ flavor invariants and flavor tensors is presented in Tab.~\ref{tab:A4 invariants and tensors}. We give some brief remarks for the mediators of each spin. 

\textbf{Scalars. }Out of the four scalar mediators, three of them, namely $\mathcal{S}_1$, $\mathcal{S}_2$, $\Xi_1$, couple to the SM bilinears of the form $\bar f^c f$, where $f=\lzv\ell,e\dzv$. Interaction terms containing these bilinears enforce a constraint on the form of the flavor tensor, due to which the tensor is allowed to be either symmetric or antisymmetric depending on the gauge structure.\footnote{This can be verified upon employing the relation of the type $\bar\psi\chi=\bar\chi^c\psi^c$. See Tab.~\ref{tab:A4 invariants and tensors} for more details and concrete examples.} Consequently, this implies that $\cS_1$ can only transform as a flavor triplet, whereby the flavor tensor needs to be antisymmetric. The interaction Lagrangians for $\cS_2$ and $\Xi_1$ scalars, on the other hand, ensure that the flavor tensor is symmetric, which allows for $\cS_2$ to transform as any of the $A_4$ singlets and $\Xi_1$ to transform either as any of the $A_4$ singlets or a triplet, where, in the latter case, the symmetric combination in the decomposition has to be selected, as indicated in Eq.~\eqref{eq:A4_tensor_product}. The remaining scalar $\varphi$ has to transform as a triplet given that it couples to the fermionic bilinear of the from $\bar\ell e$.

\textbf{Fermions. } For all of the six fermionic UV mediators, the procedure of extracting the possible flavor irreps is straightforward owing to the fact that all of the fermionic mediators couple to only one SM fermion, either $\ell$ or $e$. Therefore, depending on which SM fermion the UV mediators couples to, the UV fermionic mediator can either be any of the $A_4$ singlets or a triplet.

\textbf{Vectors. } Two out of three vector mediators, namely $\cB$ and $\cW$, can transform as any of the $A_4$ flavor irreps, since they both couple to the bilinears of the type $\bar f\gamma^\mu f$. It is also worth highlighting that $\cB$ couples to both $\bar\ell\gamma^\mu \ell$ and $\bar e \gamma^\mu e$ bilinears, which, in case of all $A_4$ singlets, allows simultaneously for two flavor tensors, as indicated in Tab.~\ref{tab:A4 invariants and tensors}. The remaining vector mediator $\cL_3$ can only transform as a triplet as it couples to the bilinear of the type $\bar e^c\gamma^\mu \ell$.


\subsection{$A_5$ flavor symmetry}
\subsubsection{Overview of the group properties}
$A_5$ symmetry group, similarly to $A_4$, is a group of even permutations defined on five objects. $A_5$ contains $5!/2=60$ elements and, in a geometric sense, $A_5$ is a symmetry group of two regular polyhedrons, more precisely dodecahedron as well as icosahedron. The defining relation expressed in terms of $S$ and $T$ generators can be written as
\begin{equation}
    S^2=T^5=(ST)^3=1.
\end{equation}

There are five irreducible representations contained in the $A_5$ group: one singlet ($\bm1$), two triplets ($\bm3$ and $\bm3'$), one quadruplet ($\bm4$) and one quintuplet ($\bm5$). As in the case of $A_4$, $S$ and $T$ generators can be specified for each irrep and the tensor products can be obtained using their explicit forms, see e.g. Refs.~\cite{Ishimori:2010au, Ding:2019xna, Novichkov:2018nkm, Feruglio:2011qq}. In the subsequent discussion, for brevity, we list only the tensor product expansions which will be relevant for the analysis and the extraction of the flavor tensors.

Starting with the $A_5$ singlet irrep, the tensor product is trivially given as
\begin{equation}\label{eq:A5 singet dec}
    \bm1\otimes\cR=\cR\otimes\bm1=\cR,
\end{equation}
where $\cR$ denotes any of the $A_5$ irreps. The tensor product of two $A_5$ triplets expands as
\begin{equation}\label{eq:A5 triplets dec}
    \bm3\otimes\bm3=\bm{1_S}\oplus\bm{3_A}\oplus\bm{5_S},
\end{equation}
where $S$ ($A$) denotes the (anti)symmetric combinations. Taking $\alpha=(\alpha_1,\alpha_2,\alpha_3)^T\sim\bm3$ and $\beta=(\beta_1,\beta_2,\beta_3)^T\sim\bm3$, the irreps in the decomposition given by Eq.~\eqref{eq:A5 triplets dec} can be written as
\begin{equation}
    \begin{alignedat}{2}
        [\alpha\beta]_{\bm{1_S}}&=\alpha_1\beta_1+\alpha_2\beta_3+\alpha_3\beta_2,
        \\
        [\alpha\beta]_{\bm{3_A}}&=
        \begin{bmatrix}
            \alpha_2\beta_3-\alpha_3\beta_2
            \\
            \alpha_1\beta_2-\alpha_2\beta_1
            \\
            \alpha_3\beta_1-\alpha_1\beta_3
        \end{bmatrix},
        \\
        [\alpha\beta]_{\bm{5_S}}&=
        \begin{bmatrix}
            2\alpha_1\beta_1-\alpha_2\beta_3-\alpha_3\beta_2
            \\
            -\sqrt{3}(\alpha_1\beta_2+\alpha_2\beta_1)
            \\
            \sqrt{6}\alpha_2\beta_2
            \\
            \sqrt{6}\alpha_3\beta_3
            \\
            -\sqrt{3}(\alpha_1\beta_3+\alpha_3\beta_1)
        \end{bmatrix}.
    \end{alignedat}
\end{equation}
Lastly, the tensor product expansion of two $A_5$ quintuplets becomes
\begin{equation}\label{eq: A5 5 dec}
    \bm5\otimes\bm5=\bm{1_S}\oplus\bm{3_A}\oplus\bm{3_A}'\oplus\bm{4_S}\oplus\bm{4_A}\oplus\bm{5_S}\oplus\bm{5_S}'.
\end{equation}
In the forthcoming analysis, only the $\bm{1_S}$ irrep from the expansion above will be required, while the remaining ones will not be showing up. Taking $\alpha\equiv(\alpha_1,\alpha_2,\alpha_3,\alpha_4,\alpha_5)^T\sim\bm5$ and $\beta\equiv(\beta_1,\beta_2,\beta_3,\beta_4,\beta_5)^T\sim\bm5$, we get
\begin{equation}
    [\alpha\beta]_{\bm{1_S}}=\alpha_1\beta_1+\alpha_2\beta_5+\alpha_3\beta_4+\alpha_4\beta_3+\alpha_5\beta_2.
\end{equation}
Details regarding the remaining tensor product decompositions written out in the component form can be found in, e.g. Ref.~\cite{Ding:2019xna}.

\subsubsection{Extraction of flavor tensors}
Analogously to the case of $A_4$ flavor symmetry, in addition to being motivated by the similar choice in the various $A_5$ models~\cite{Feruglio:2011qq, Ding:2019xna, Puyam:2023div}, the SM leptons are taken to transform as\footnote{It is straightforward to show that if we opt for the arrangement in which $\ell\sim\bm3'$, we obtain the same flavor tensors and, consequently, same SMEFT matching relations. Also notice that under this particular arrangement of leptons under the $A_5$ flavor symmetry, the quadruplet ($\bm4$) irrep cannot appear in the decompositions.}
\begin{equation}
    \ell\equiv(\ell_1,\ell_2,\ell_3)^T\sim\bm3,\quad (e_1,e_2,e_3)\sim(\bm1,\bm1,\bm1).
\end{equation}

In light of this arrangement, along with the decompositions given by Eqs.~\eqref{eq:A5 singet dec}, \eqref{eq:A5 triplets dec} and \eqref{eq: A5 5 dec} we can perform the analogous analysis as in the $A_4$ case and extract the possible $A_5$ irreps for the UV mediators and determine the form of the flavor tensors. The complete classification of the UV mediators, complemented by the $A_5$ invariants and the flavor tensors extracted by demanding the interaction Lagrangian be invariant under the $A_5$ flavor symmetry is presented in Tab.~\ref{tab:A5 invariants and tensors}. A few brief remarks for the UV mediators are in order.

\textbf{Scalars. }Due to the antisymmetric nature of the $\cS_1$ flavor tensor, in order to make the Lagrangian invariant under the $A_5$ flavor symmetry, the only irrep we can associate to this field is a triplet representation. $\cS_2$, on the other hand, has to transform as a singlet, whereby we use the fact that the flavor tensor has to be symmetric to relate some of the independent parameters in the flavor tensor. $\varphi$ scalar, coupling to the $\bar\ell e\sim\bm3$ bilinear, has to transform as a triplet accordingly. Lastly, $\Xi_1$, with its symmetric flavor tensor, can either transform as a singlet or a quintuplet, while the triplet irrep is not allowed because of the antisymmetric combination, as indicated in Eq.~\eqref{eq:A5 triplets dec}.

\textbf{Fermions. }As demonstrated in the previous case of $A_4$ symmetry, flavor irreps for the fermionic mediators are directly inherited from the SM fermions they are being coupled to. Consequently, fermionic mediators can either transform as singlets or triplets.

\textbf{Vectors. }$\cB$ and $\cW$ can both transform either as a singlet, triplet or quintuplet, where in case of $\cB$ only the singlet irrep allows for both flavor tensors simultaneously. Once again, $\cL_1$ vector can only transform as a triplet. 

\subsection{$S_4$ flavor symmetry}

\subsubsection{Overview of the group properties}
The group $S_4$ represents a finite, permutation group of four distinct objects containing $4!=24$ elements. Geometrically, $S_4$ is isomorphic to the symmetry group of a regular octahedron. The defining relation for $S$ and $T$ generators in this case is given as
\begin{equation}
    S^2=T^4=(ST)^3=1,
\end{equation}
where the explicit form of the $S$ and $T$ generator can be found, e.g. in Refs~\cite{Ishimori:2010au, Bazzocchi:2012st, Krishnan:2012me}.

Similarly to the $A_5$ group considered previously, the group $S_4$ also has five irreducible representations: two singlets ($\bm1$ and $\bm1'$), one doublet ($\bm2$) and two triplets ($\bm3$ and $\bm3'$).

When it comes to the tensor product decompositions, starting once again with the $S_4$ singlets, we have the products of the form
\begin{equation}\label{eq:S4 singet dec}
    \bm1\otimes\cR=\cR\otimes\bm1=\cR,\quad \bm1'\otimes\bm1'=\bm1,
\end{equation}
where $\cR$ denotes any of the $S_4$ irreps. The tensor product of the two $S_4$ triplets of the same type becomes
\begin{equation}\label{eq:S4 triplet dec}
    \bm3\otimes\bm3=\bm3'\otimes\bm3'=\bm1\oplus\bm2\oplus\bm{3_S}\oplus\bm{3_A}.
\end{equation}
Taking $\alpha\equiv(\alpha_1,\alpha_2,\alpha_3)^T\sim\bm3$ and $\beta\equiv(\beta_1,\beta_2,\beta_3)^T\sim\bm3$, we have
\begin{equation}
    \begin{alignedat}{2}
        [\alpha\beta]_{\bm{1_S}}&=\alpha_1\beta_1+\alpha_2\beta_2+\alpha_3\beta_3,
        \\
        [\alpha\beta]_{\bm{2}}&=
        \begin{bmatrix}
            \frac{1}{\sqrt2}(\alpha_2\beta_2-\alpha_3\beta_3)
            \\
            \frac{1}{\sqrt6}(-2\alpha_1\beta_1+\alpha_2\beta_2+\alpha_3\beta_3)
        \end{bmatrix},
        \\
        [\alpha\beta]_{\bm{3_S}}&=
        \begin{bmatrix}
            \alpha_2\beta_3+\alpha_3\beta_2
            \\
            \alpha_1\beta_3+\alpha_3\beta_1
            \\
            \alpha_1\beta_2+\alpha_2\beta_1
        \end{bmatrix},
        \quad 
        [\alpha\beta]_{\bm{3_A}}=
        \begin{bmatrix}
            \alpha_3\beta_2-\alpha_2\beta_3
            \\
            \alpha_1\beta_3-\alpha_3\beta_1
            \\
            \alpha_2\beta_1-\alpha_1\beta_2
        \end{bmatrix}.
    \end{alignedat}
\end{equation}
From the product of two $S_4$ doublets, which decomposes as
\begin{equation}\label{eq:S4 doublet dec}
    \bm2\otimes\bm2=\bm{1}\oplus\bm{1}'\oplus\bm2,
\end{equation}
only the trivial singlet $\bm1$ will be required in the following discussion, which, written out in terms of components taking $\alpha\equiv(\alpha_1,\alpha_2)^T\sim\bm2$ and $\beta\equiv(\beta_1,\beta_2)^T\sim\bm2$, becomes
\begin{equation}
    [\alpha\beta]_{\bm{1}}=\alpha_1\beta_1+\alpha_2\beta_2.
\end{equation}
Remaining tensor product decompositions can be found in Refs.~\cite{Ishimori:2010au, Bazzocchi:2012st, Hagedorn:2006ug, Izawa:2023qay}.

\subsubsection{Extraction of flavor tensors}
Analogously to the previous two sections, under the $S_4$ flavor symmetry, we take the SM lepton fields to transform as~\cite{Krishnan:2012me}\footnote{Other configurations, which can be found in the literature, involve constructing the right-handed triplet $e\equiv(e_1,e_2,e_3)^T\sim\bm3$, as in e.g. Ref.~\cite{Hagedorn:2006ug}, or grouping the first and second generation into a doublet of the form $(e_1,e_2)^T\sim\bm2$ and leaving the third generation as a singlet $e_3\sim\bm1$, as in Ref.~\cite{Izawa:2023qay}.}
\begin{equation}
    \ell\equiv(\ell_1,\ell_2,\ell_3)^T\sim\bm3,\quad (e_1,e_2,e_3)\sim(\bm1,\bm1,\bm1).
\end{equation}
Following the same approach as in the previous two cases, using the Eqs.~\eqref{eq:S4 singet dec}, \eqref{eq:S4 triplet dec} and \eqref{eq:S4 doublet dec}, we perform the classification of the NP mediators and extract the flavor tensors. We present these results in Tab.~\ref{tab:S4 invariants and tensors} and offer a few brief comments on the procedure.

\textbf{Scalars. }Starting with $\cS_1$ scalar, we once again notice that in the presence of the $S_4$ flavor symmetry this scalar has to transform as a triplet, in order to be able to isolate the antisymmetric combination from the product decomposition. $\cS_2$ and $\varphi$ can only transform as a singlet and triplet, respectively, while the allowed flavor irreps for $\Xi_1$ turn out to be singlet, doublet and triplet representations, since for all of them the symmetric combination exists in the tensor product decomposition.

\textbf{Fermions. }As in the previous two cases, NP fermions inherently acquire the flavor structure from the SM lepton they are coupled to, which causes them to transform either as singlets or triplets.

\textbf{Vectors. }$\cB$ and $\cW$ vectors are allowed to transform as singlets, doublets and triplets under the $S_4$ symmetry, where, in the case of $\cB$, the singlet flavor irrep allows for both interaction terms, as indicated in Tab.~\ref{tab:S4 invariants and tensors}. $\cL_3$, conversely, is only allowed to transform as a triplet, on account of its interaction Lagrangian.

\section{Discrete leptonic directions and phenomenology}
\label{sec:directions_and_pheno}

\subsection{SMEFT Matching}

Following the discussion of the properties of the $A_4$, $A_5$ and $S_4$ flavor groups, which facilitated the extraction of flavor tensors and classification of the NP mediators presented in Tabs.~\ref{tab:A4 invariants and tensors}, \ref{tab:A5 invariants and tensors} and \ref{tab:S4 invariants and tensors}, in this section we explore the implications of the aforementioned flavor symmetries once the NP mediators are integrated out at tree level and matched onto the dimension-6 SMEFT operators. In this phase, we note, following the earlier analysis in Ref.~\cite{Greljo:2023adz}, that the matching is performed taking a single NP mediator at a time. In order to find the tree-level dimension-6 matching relations, we take advantage of the generic matching results derived in absence of any flavor symmetries~\cite{deBlas:2017xtg}. Using the flavor tensors derived in the previous section, the SMEFT matching Lagrangians for each NP flavor irrep can be readily obtained.\footnote{As indicated in Ref.~\cite{deBlas:2017xtg}, the expressions for various SMEFT Wilson coefficients are obtained upon evaluating the sum over flavor indices of the NP mediator. This sum, in case of a singlet flavor irrep, trivially reduces to a single term, while the full sum has to be evaluated for all other flavor irreps~\cite{Greljo:2023adz}.} The matching results for different flavor assumptions and for those irreps that match onto the discrete leptonic SMEFT directions (see Sec.~\ref{sec:setup}) are collected in Tabs.~\ref{tab:A4 SMEFT Matching}, \ref{tab:A5 SMEFT Matching} and \ref{tab:S4 SMEFT Matching}. We concisely discuss the results for NP mediators of each spin.

\begin{table}[t]
\centering
\begin{tabular}{ccccc}
\toprule
\textbf{Label} & \textbf{Operator}& \textbf{Label}& \textbf{Operator}
\\
\midrule
$[\cO_{\ell\ell}]_{ijkl}$
&$(\bar\ell_i\gamma^\mu \ell_j)(\bar\ell_k\gamma_\mu\ell_l)$
&$[\cO_{\phi\ell}^{(1)}]_{ij}$
&$(\phi^\dag i\overset{\text{\footnotesize$\leftrightarrow$}}{D}_\mu \phi)(\bar \ell_i\gamma^\mu \ell_j)$
\\[3pt]
$[\cO_{\phi e}]_{ij}$
&$(\phi^\dag i\overset{\text{\footnotesize$\leftrightarrow$}}{D}_\mu \phi)(\bar e_i\gamma^\mu e_j)$
&$[\cO_{\phi\ell}^{(3)}]_{ij}$
&$(\phi^\dag i\overset{\text{\footnotesize$\leftrightarrow$}}{D^a_\mu} \phi)(\bar \ell_i\gamma^\mu \sigma^a \ell_j)$
\\
\bottomrule
\end{tabular}
\caption{Dimension-6 SMEFT operators emerging in the SMEFT matching relations in case of discrete leptonic directions. See Tabs.~\ref{tab:A4 SMEFT Matching}, \ref{tab:A5 SMEFT Matching} and \ref{tab:S4 SMEFT Matching} for the SMEFT matching relations.}
\label{tab:SMEFToperators}
\end{table}

\textbf{Scalars. }For all flavor assumptions considered in this analysis, $\cS_1$ and $\Xi_1$ scalars match onto a leptonic direction regardless of their flavor irrep. This arises from the fact that for each flavor irrep there is only one allowed flavor invariant, which consequently leads to the flavor tensor depending on a single parameter. $\cS_2$ and $\varphi$ scalars don't generate a direction since there are more allowed flavor invariants, which, without further assumptions, come with different independent parameters. In the forthcoming phenomenological analysis, such cases will not be treated.

\textbf{Fermions. }Starting with $A_4$ flavor symmetry, flavor irreps of all NP fermions generate a SMEFT direction once integrated out at tree level. The situation is different in case of $A_5$ and $S_4$ flavor assumptions, where only flavor triplets generate a direction, while the matching condition for singlet irreps always comes with three independent parameters.

\textbf{Vectors. }For all three flavor assumptions, the singlet irreps in case of $\cB$ allow for both interaction terms with more than one independent parameter, which, as a result, leads to the SMEFT matching condition containing more than one parameter. All other flavor irreps allow for only one interaction term and one flavor invariant leading to the SMEFT direction in the matching procedure. Completely analogous reasoning follows for $\cW$ vector, which for all flavor irreps yields a leptonic direction. On the other hand, $\cL_3$ vector never matches onto a leptonic direction since there are always three parameters present in the flavor tensor.

\begin{figure*}[t]
\begin{minipage}[t]{\textwidth}
\makebox[\linewidth]{
  \includegraphics[width=1.0\linewidth]{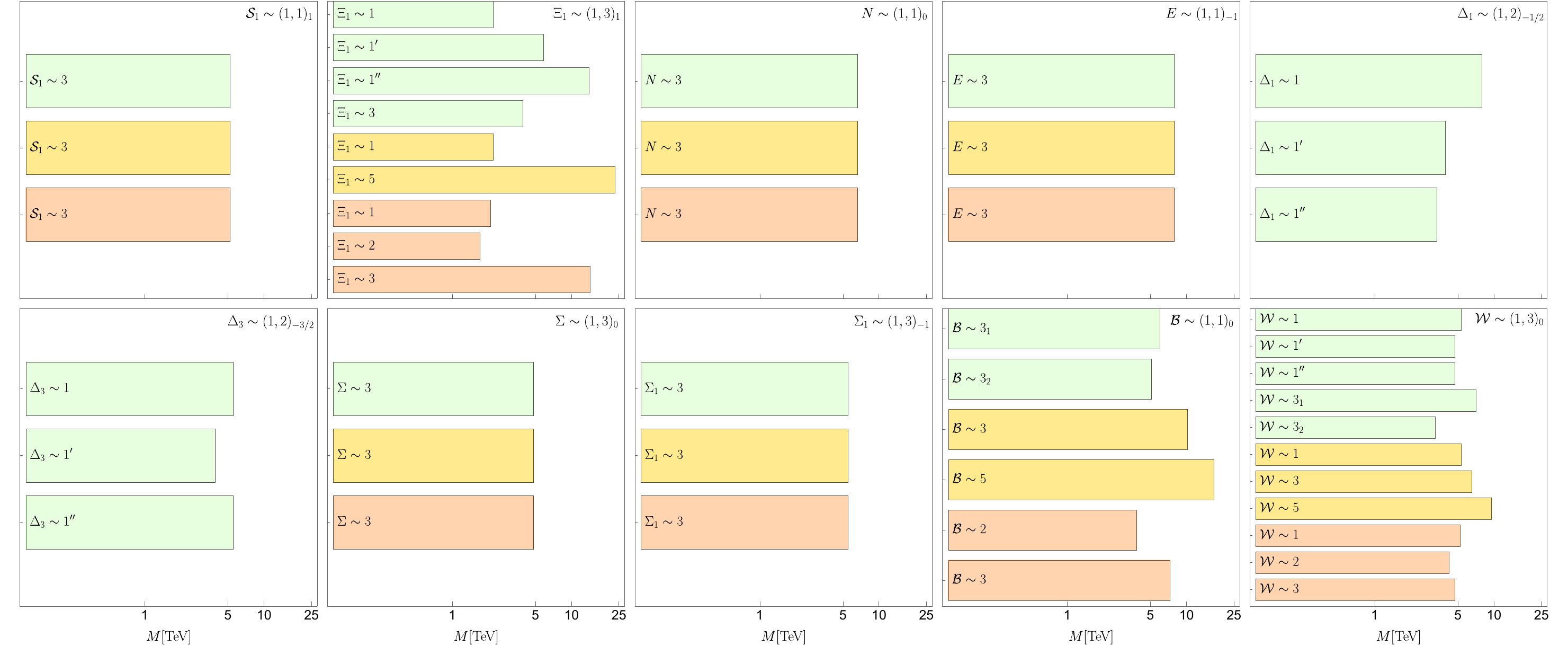}}
\end{minipage}
\caption{Overview of the tree-level bounds on the mass scale for various irreps using low-energy observables. Green bars indicate $A_4$, yellow $A_5$ and orange $S_4$ flavor irreps. See Sec.~\ref{sec: low-energy obs} for more details and Tabs.~\ref{tab:A4 SMEFT Matching}, \ref{tab:A5 SMEFT Matching} and \ref{tab:S4 SMEFT Matching} for SMEFT matching relations.}
\label{fig: low-energy obs}
\end{figure*}

\subsection{Phenomenology}
\label{sec: pheno}

As the concluding segment of our analysis, we carry out the phenomenological examination of the discrete leptonic directions presented in Tabs.~\ref{tab:A4 SMEFT Matching}, \ref{tab:A5 SMEFT Matching} and \ref{tab:S4 SMEFT Matching}. As stated previously, directions are well-defined linear combinations of SMEFT operators multiplied by a single parameter, allowing for a phenomenological analysis through which the lower bounds on the mass scale for various flavor irreps can be obtained.\footnote{Since the parameter entering the SMEFT directions is of the form $g^2/M^2$, where $g$ and $M$ are coupling and mass respectively, the phenomenological analysis is conducted by setting the coupling to unity and obtaining the lower bound on the scale $M$.} With our attention directed towards the leptonic sector in this analysis, we focus on the low-energy observables and the constraints originating from the transitions involving charged lepton flavor violation (cLFV).

\subsubsection{Low-energy observables}
\label{sec: low-energy obs}

In order to determine the bounds on the mass scales using the low-energy observables, we employ the fits presented in Refs.~\cite{Falkowski:2017pss, Breso-Pla:2023tnz, Falkowski:2015krw}. Given that the discrete leptonic directions are constituted of SMEFT operators involving four leptons only, the datasets of particular relevance involve lepton pair production in $e^+e^-$ collisions~\cite{ALEPH:2013dgf, VENUS:1997cjg}, $W/Z$ pole observables~\cite{Falkowski:2019hvp, Efrati:2015eaa} in addition to $\nu_\mu e$ scattering~\cite{ParticleDataGroup:2016lqr}, $\tau$ polarization~\cite{VENUS:1997cjg}, low-energy parity violating $e^+e^-$ scattering and neutrino trident production~\cite{CHARM-II:1990dvf, CCFR:1991lpl, Altmannshofer:2014pba}. The bounds on the scale for various NP flavor irreps, which integrate out to SMEFT direction, are presented in Fig.~\ref{fig: low-energy obs}. A few observations are in order.

\textbf{Scalars. }$\cS_1$ and $\Xi_1$ scalars generate leptonic directions for all flavor assumptions. Starting with $\cS_1$, as discussed previously, it has to transform as a flavor triplet under the flavor assumptions considered in this work, which consequently leads to the same form of flavor tensor and same bounds from low-energy observables of $\cO(5\text{ TeV})$. All flavor irreps of $\Xi_1$ have lower bound which is greater than 1 TeV and in a few particular cases, such as $\bm1''$ ($A_4$), $\bm5$ ($A_5$) and $\bm3$ ($S_4$), the lower bound is found to be greater than 10 TeV. 

In case of $\cS_1$ scalar, the observables from which the lower bounds on all of the flavor irreps are extracted involve $e^+e^-\to \mu^+\mu^-$ forward ($\sigma_F$) and backward ($\sigma_B$) scattering, low-energy neutrino scattering ($\nu_\mu e^-\to\nu_\mu e^-$), leptonic tau decay $\tau^-\to \mu^- \nu_\tau\bar\nu_\mu$ as well as the $\delta m_W$~\cite{Falkowski:2015krw}. From the differential cross section of the Bhabha scattering ($e^+e^-\to e^+e^-$), parity-violating electron scattering as well as the tau decay $\tau^-\to \mu^-\nu_\mu\bar\nu_\tau$, the bounds on the $\Xi_1\sim\bm1$ ($A_4$) and $\Xi_1\sim\bm1$ ($A_5$) are derived, whereas in case of $\Xi_1\sim\bm1$ ($S_4$) and $\Xi_1\sim\bm2$ ($S_4$) only the former two observables are relevant. The leptonic tau decay $\tau^-\to e^-\nu_e\bar\nu_\tau$ is relevant in the extraction of the lower bound for $\Xi_1\sim\bm1'$ ($A_4$) irrep, while the $\Xi_1\sim\bm1''$ ($A_4$) irrep receives the bound predominantly from the $e^+e^-\to \mu^+\mu^-$ forward ($\sigma_F$) and backward ($\sigma_B$) scattering and low-energy neutrino scattering ($\nu_\mu e^-\to\nu_\mu e^-$). The remaining flavor irreps are bounded taking the combined fit of all of the aforementioned observables.

\textbf{Fermions. }$N$, $E$, $\Sigma$ and $\Sigma_1$ fermions all transform as flavor triplets under the three flavor assumptions. For these four fermions, all flavor irreps yield the same lower bound on the mass scale due to the identical form of the flavor tensor and, consequently, the SMEFT matching relations. As can be read off from Fig.~\ref{fig: low-energy obs}, the lower bounds for all irreps turn out to be of $\cO(5\text{ TeV})$. In comparison, $\Delta_1$ and $\Delta_3$ fermions only match onto directions in case of $A_4$ flavor symmetry and the bounds on their flavor irreps are found to be ranging from 3 TeV up to 8 TeV.

The lower bounds on all of the flavor irreps of $N$, $E$, $\Sigma$ and $\Sigma_1$ fermions can be extracted from the combining fit involving low-energy neutrino scattering ($\nu_\mu e^-\to\nu_\mu e^-$), leptonic tau decays, $\delta m_W$ and parity-violating electron scattering, as the $[\cO_{\phi\ell}^{(1)}]_{ij}$ and $[\cO_{\phi\ell}^{(3)}]_{ij}$ SMEFT operators present in the matching relations for these flavor irreps enter these observables~\cite{Falkowski:2015krw}. Conversely, as $\Delta_1$ and $\Delta_3$ fermions match onto the $[\cO_{\phi e}]_{ij}$ SMEFT operator, the lower bounds on their flavor irreps can be extracted using the combined fit of low-energy neutrino scattering ($\nu_\mu e^-\to\nu_\mu e^-$) as well as parity-violating electron scattering.

\textbf{Vectors. }Most of the flavor irreps of $\cB$ and $\cW$ vectors have the lower bound of $\cO(5\text{ TeV})$, except for a few specific cases, e.g. $\cB\sim\bm3$ ($A_5$), $\cB\sim\bm5$ ($A_5$) and $\cW\sim\bm5$ ($A_5$), whose bounds are of $\cO(10\text{ TeV})$. 

Regarding the relevant observables which can be used to determine the lower bounds on the flavor irreps of the vector mediators, due to the fact that all of the irreps match onto the $[\cO_{\ell\ell}]_{ijkl}$ SMEFT operator, which is also the case for the scalar mediators, this implies that the similar discussion as in the case of scalar mediators can be applied in this case and that the similar set of observables will be relevant in the process of determining the lower bounds for the vector flavor irreps.

\subsubsection{Charged lepton flavor violation (cLFV)}
\label{sec: cLFV obs}
A distinctive new feature surfacing from imposing discrete flavor symmetries at the level of UV lagrangians describing the interactions of NP mediators with the SM fields along with integrating them out at tree level, is the emergence of the SMEFT operators in the matching relations which contribute to $|\Delta L_\alpha|=1$ and $|\Delta L_\alpha|=2$ transitions, where $\alpha=e,\mu,\tau$. Contrary to the similar analysis in presence of continuous flavor symmetries, where this class of operators would emerge with significant flavor suppression~\cite{Greljo:2022cah}, in case of discrete flavor symmetries, these operators appear at the same order in flavor power counting. As a result, this opens up a new phenomenological sector that can be used to put bounds on the relevant NP irreps~\cite{Heeck:2024uiz, Heeck:2016xwg, Conlin:2020veq, Banerjee:2022vdd, Bigaran:2022giz, Crivellin:2013hpa, Hayasaka:2010np}. 

\begin{figure}[t]
    \centering
    \includegraphics[width=1.0\linewidth]{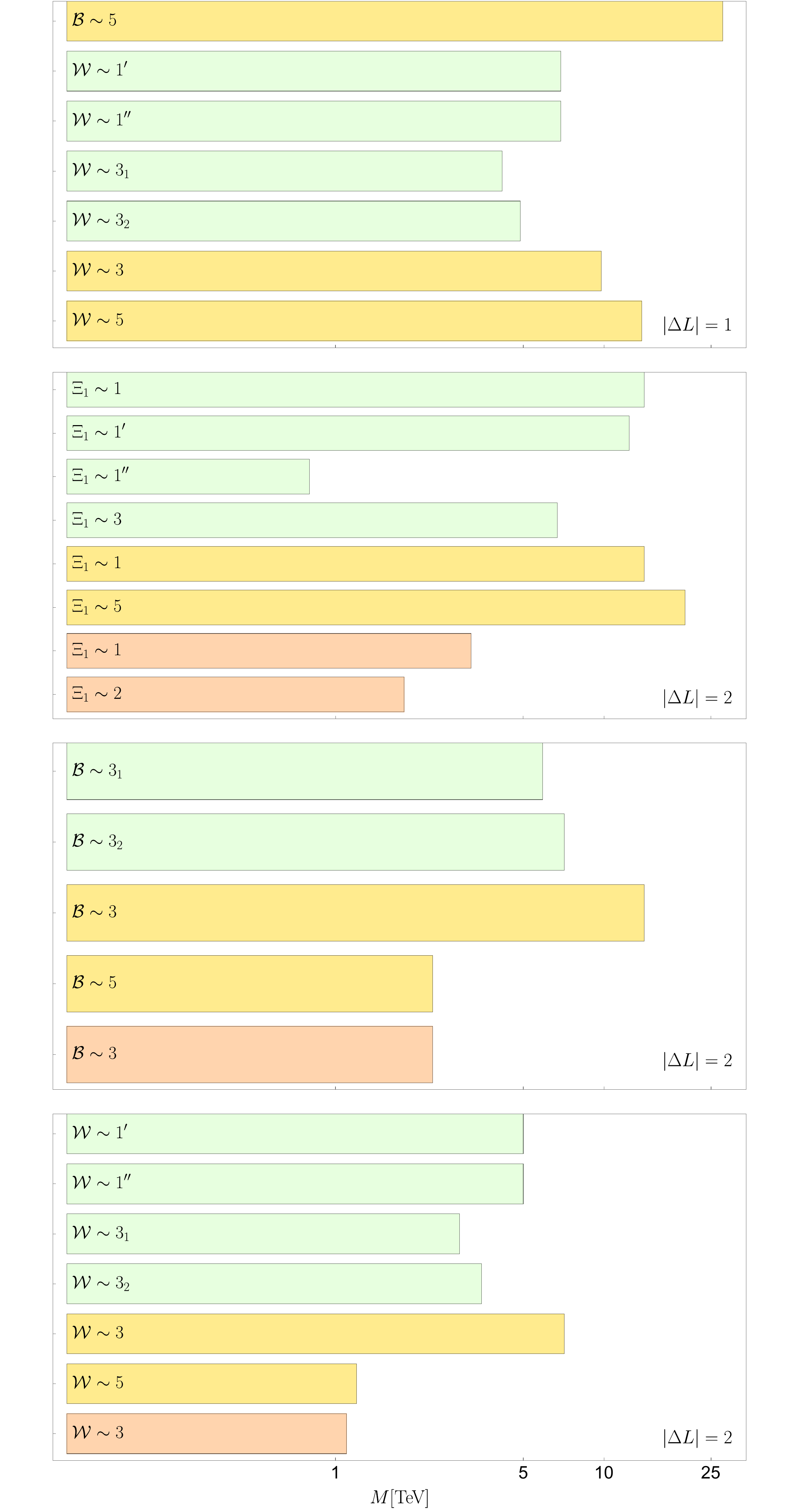}
    \caption{Overview of the tree-level bounds on the scale for various irreps using upper limits on the $|\Delta L_\alpha|=1$ (the uppermost plot) and $|\Delta L_\alpha|=2$ (remaining three plots) cLFV SMEFT operators. Green bars indicate $A_4$, yellow $A_5$ and orange $S_4$ flavor irreps. See Sec.~\ref{sec: cLFV obs} for more details and Tabs.~\ref{tab:A4 SMEFT Matching}, \ref{tab:A5 SMEFT Matching} and \ref{tab:S4 SMEFT Matching} for SMEFT matching relations.}
    \label{fig: CLFV12}
\end{figure}

Starting with the tree-level $|\Delta L_\alpha|=1$ operators, the relevant transitions for estimating the bounds on the mass scale of the relevant irreps are $\tau^\pm\to e^\pm\mu^+\mu^-$ and $\tau^\pm\to \mu^\pm e^+ e^-$. For these two transitions, the branching ratios expressed in terms of the Wilson coefficients appearing in the discrete leptonic directions can approximately be expressed as~\cite{Crivellin:2013hpa} 
\begin{equation}\label{eq: cLFV1 BR}
    \begin{alignedat}{2}
        \text{Br}(\tau^\pm\to e^\pm\mu^+\mu^-)&\approx\frac{1}{\Gamma_\tau}\frac{m_\tau^5}{1536\pi^3}\left|[\cC_{\ell\ell}]_{1322}\right|^2,
        \\
        \text{Br}(\tau^\pm\to \mu^\pm e^+ e^-)&\approx\frac{1}{\Gamma_\tau}\frac{m_\tau^5}{1536\pi^3}\left|[\cC_{\ell\ell}]_{2311}\right|^2,
    \end{alignedat}
\end{equation}
where $m_\tau$ and $\Gamma_\tau$ are the mass and total decay width of the tau, respectively. Upon reviewing the Tabs.~\ref{tab:A4 SMEFT Matching} and \ref{tab:A5 SMEFT Matching}, it can be concluded that the Wilson coefficients entering the Eq.~\eqref{eq: cLFV1 BR} are found in SMEFT matching relations for $A_4$ and $A_5$ flavor irreps of $\cB$ and $\cW$ vector mediators. Using the values for the branching ratios~\cite{ParticleDataGroup:2024cfk}
\begin{equation}
    \begin{alignedat}{2}
        \text{Br}(\tau^\pm\to e^\pm\mu^+\mu^-)&<2.7\times10^{\eminus8},
        \\
        \text{Br}(\tau^\pm\to \mu^\pm e^+ e^-)&<1.8\times10^{\eminus8},
    \end{alignedat}
\end{equation}
which are readily translated to the bounds on the corresponding Wilson coefficients
\begin{equation}
    |[\cC_{\ell\ell}]_{1322}|<(8.84\text{ TeV})^{\eminus2},
    \quad 
    |[\cC_{\ell\ell}]_{2311}|<(9.76\text{ TeV})^{\eminus2},
\end{equation}
we can obtain the lower bounds on the masses for the different irreps. The bounds obtained from these transitions are collected in the topmost plot in Fig.~\ref{fig: CLFV12}. 
In comparison to the low-energy observables, we conclude that the bounds are of similar magnitude, apart from the $\cB\sim\bm5$ ($A_5$) and $\cW\sim\bm3$ ($A_5$) irreps, for which the bounds obtained from $|\Delta L_\alpha|=1$ transition are improved by a factor of $\sim2$.

Similarly, the tree-level $|\Delta L_\alpha|=2$ operators appear in the SMEFT matching once the various flavor irreps of $\Xi_1$ scalar and $\cB$ and $\cW$ vectors are integrated out (see Tabs.~\ref{tab:A4 SMEFT Matching}, \ref{tab:A5 SMEFT Matching} and \ref{tab:S4 SMEFT Matching}). As outlined in Ref.~\cite{Heeck:2024uiz}, $|\Delta L_\alpha|=2$ SMEFT operators with the best upper bounds entering the leptonic directions in case of $\Xi_1$, $\cB$ and $\cW$ mediators are listed below:
\begin{itemize}
    \item $[\cC_{\ell\ell}]_{2121}$, corresponding to $\Delta L_\mu=-\Delta L_e=2$ transition with upper bound of $(3.2\text{ TeV})^{\eminus2}$ obtained from the $\text{Mu}-\text{to}-\overline{\text{Mu}}$ transition from the PSI experiment~\cite{Willmann:1998gd}.
    \item $[\cC_{\ell\ell}]_{2131}$, which corresponds to $\Delta L_e=-2\Delta L_\tau=-2\Delta L_\mu =2$ transition with upper bound of $(10.0\text{ TeV})^{\eminus2}$ obtained from $\tau^\pm\to\mu^\mp e^\pm e^\pm$~\cite{Hayasaka:2010np}.
    \item $[\cC_{\ell\ell}]_{1232}$, which corresponds to $\Delta L_\mu=-2\Delta L_\tau=-2\Delta L_e =2$ transition with upper bound of $(8.8\text{ TeV})^{\eminus2}$ obtained from $\tau^\pm\to e^\mp \mu^\pm\mu^\pm$~\cite{Hayasaka:2010np}.
\end{itemize}

These bounds on the relevant $|\Delta L_\alpha|=2$ operators can once again be directly translated to the lower bound on the scales of various irreps. We collect the mass bounds for various irreps of $\Xi_1$, $\cB$ and $\cW$ mediators in the lower three plots in Fig.~\ref{fig: CLFV12}. In comparison to the bounds obtained using the low-energy observables, it can be observed that in some distinct cases, bounds extracted from the cLFV operators are yet again improved by a factor of $\sim2$, e.g. $\Xi_1\sim\bm1$ ($A_4$), $\Xi_1\sim\bm1'$ ($A_4$) and $\Xi_1\sim\bm1$ ($A_5$). Alongside this, comparing the bounds derived from the $|\Delta L_\alpha|=1$ to the $|\Delta L_\alpha|=2$ transitions for the $\cB$ and $\cW$ flavor irreps, it becomes apparent that $|\Delta L_\alpha|=1$ bounds are stronger in most cases.

\begin{figure}[t]
    \centering
    \includegraphics[width=1.0\linewidth]{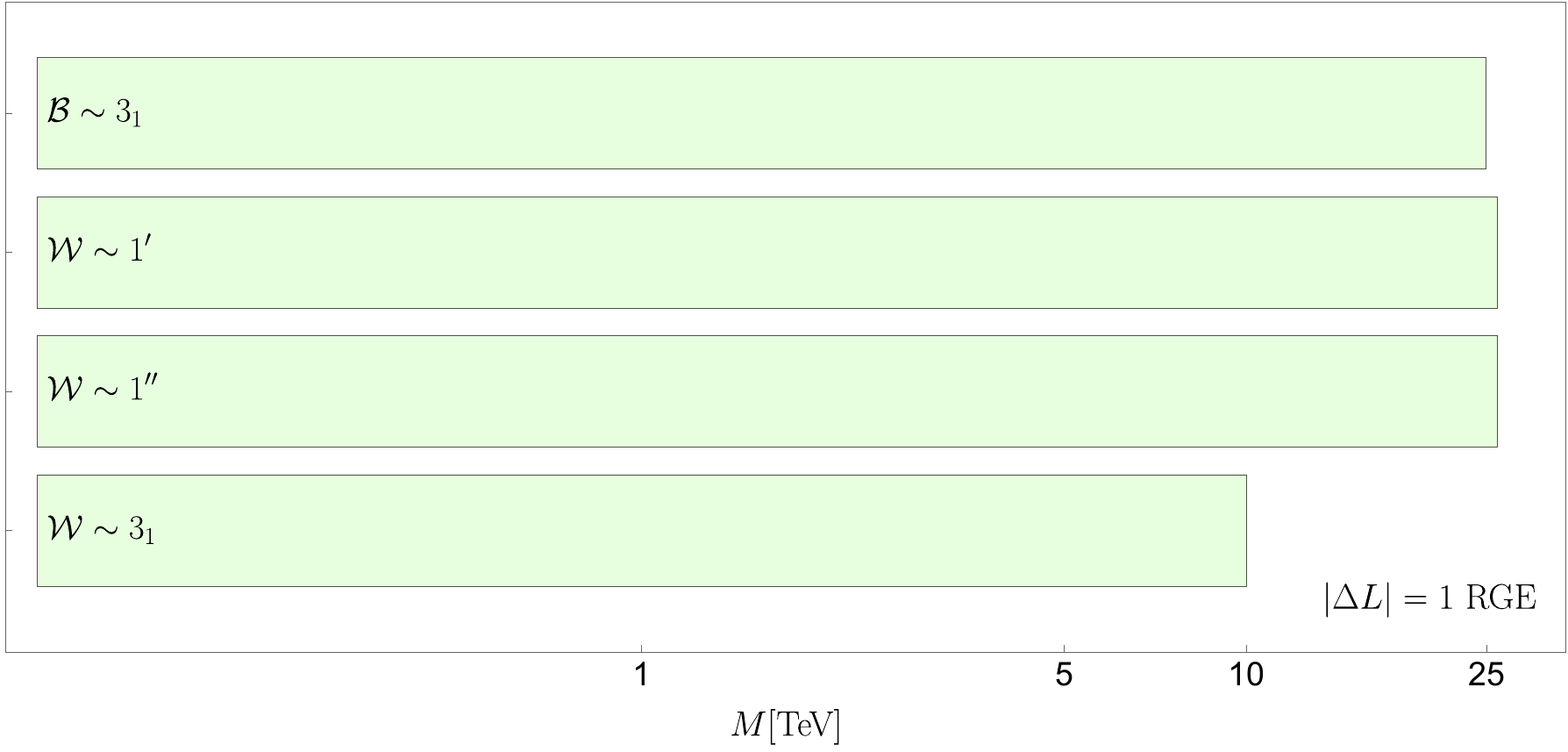}
    \caption{Overview of the bounds on the mass scale for various $A_4$ flavor irreps emerging from the RG-generated $|\Delta L_\alpha|=1$ transitions in the leading-log (LL) approximation. The initial and final running scales are taken to be $\mu_i\sim3\text{ TeV}$ and $\mu_f\sim100\text{ GeV}$~\cite{Greljo:2023bdy}. See Sec.~\ref{sec: cLFV obs} for more details and Tab.~\ref{tab:A4 SMEFT Matching} for SMEFT matching relations.}
    \label{fig: CLFVmu3e}
\end{figure}

As the final component of our phenomenological analysis, we dedicate our attention to the bounds on the four-lepton operators originating from the $|\Delta L_\alpha|=1$ $\mu^\pm\to e^\pm e^+ e^- $ and $\mu^\pm\to e^\pm\gamma$ transitions. Within the context of the discrete leptonic directions considered in this analysis, the procedure of translating the bounds on the relevant irreps, however, necessitates moving beyond the tree-level analysis and focusing on the RGE effects, as the SMEFT operators entering these transitions are not generated at tree level (see Tabs.~\ref{tab:A4 SMEFT Matching}, \ref{tab:A5 SMEFT Matching} and \ref{tab:S4 SMEFT Matching}). Combining the bounds from these two transitions enables the extraction of bounds on the four-lepton operators, which are given as~\cite{Davidson:2020hkf, Ardu:2024bua}
\begin{equation}\label{eq26}
    |\cC^{ee}_{V,LL}|\leq 7.0\times 10^{\eminus7},
    \qquad 
    |\cC^{ee}_{V,LR}|\leq 1.0\times 10^{\eminus6}.
\end{equation}
Furthermore, in the leading-log (LL) approximation, the coefficients $\cC^{ee}_{V,LL}$ and $\cC^{ee}_{V,LR}$ can be estimated as~\cite{Alonso:2013hga}
\begin{equation}\label{eq27}
    \begin{alignedat}{2}
        \cC^{ee}_{V,LL}&=\frac{v^2}{16\pi^2}\lzs\frac{2g_1^2}{3}[\cC_{\ell\ell}]_{2133}+\frac{g_1^2+g_2^2}{3}[\cC_{\ell\ell}]_{2331}\dzs\log\frac{\mu_f}{\mu_i},
        \\
        \cC^{ee}_{V,LR}&=\frac{v^2}{16\pi^2}\lzs \frac{8g_1^2}{3}[\cC_{\ell\ell}]_{2133}+\frac{4g_1^2}{3}[\cC_{\ell\ell}]_{2331} \dzs\log\frac{\mu_f}{\mu_i},
    \end{alignedat}
\end{equation}
where $v=246\text{ GeV}$, $g_1$ and $g_2$ are the SM gauge coupling constants and $\mu_i$ ($\mu_f$) are the initial (final) running scale.\footnote{The contributions to the RGEs proportional to the Yukawa matrices have been neglected, as we take $[Y_e]_{rs}\approx y_\tau\delta_{r3}\delta_{s3}$~\cite{Jenkins:2013wua}.} Synthesizing the Eqs.~\eqref{eq26} and \eqref{eq27} together with the SMEFT matching results presented in Tabs.~\ref{tab:A4 SMEFT Matching}, \ref{tab:A5 SMEFT Matching} and \ref{tab:S4 SMEFT Matching} leads to the conclusion that this effect is particularly relevant for bounds on certain $A_4$ irreps, which are collected in Fig.~\ref{fig: CLFVmu3e}. Despite the fact that these bounds are generated beyond tree level, they are found to be significantly stronger compared to the corresponding tree-level ones (see Fig.~\ref{fig: low-energy obs} and \ref{fig: CLFV12} for more details). 

\section{Conclusion and outlook}
\label{sec:conclusions}

The enigmatic structure of the lepton sector encourages the thorough examination of a range of distinct flavor groups aiming to portray the peculiar mass and mixing textures. Deepening the exploration of the previous findings on the interplay of the flavor symmetries and the UV mediators culminating in comprehensive phenomenological analyses~\cite{Greljo:2022cah, Greljo:2023adz, Greljo:2023bdy}, this work addresses three well-motivated flavor symmetries, more specifically $A_4$, $A_5$ and $S_4$, in the context of NP mediators coupling to SM lepton fields.

Examining the fundamental properties of these discrete flavor symmetry groups, the complete classification of the NP mediators in terms of the possible flavor irreps is performed and the flavor structure of the corresponding coupling matrices is determined. These results are collected in Tabs.~\ref{tab:A4 invariants and tensors}, \ref{tab:A5 invariants and tensors} and \ref{tab:S4 invariants and tensors}. In the next step, utilizing the results outlined in Ref.~\cite{deBlas:2017xtg}, taking into account the assumption that SM is extended by a single NP mediator at a time, we integrate out various NP flavor irreps at tree level and obtain the matching onto dimension-6 SMEFT operators. Primary emphasis is given to the SMEFT \textit{discrete leptonic directions}, i.e. well-defined linear combinations of dimension-6 SMEFT operators multiplied by a single parameter. These matching relations are summarized in Tabs.~\ref{tab:A4 SMEFT Matching}, \ref{tab:A5 SMEFT Matching} and \ref{tab:S4 SMEFT Matching}.

Owing to their reliance on a single parameter, these matching relations can be employed in the phenomenological analyses, whose primary goal is to estimate the lower bound on the mass scale of a given NP flavor irrep. The analysis presented in Sec.~\ref{sec: pheno} is segmented into two parts: the first addresses the derivation of the bounds using the low-energy observables, while the second examines the $|\Delta L_\alpha|=1$ and $|\Delta L_\alpha|=2$ cLFV transitions. A thorough compilation of these bounds is presented in Fig.~\ref{fig: low-energy obs}, \ref{fig: CLFV12} and \ref{fig: CLFVmu3e}. For the pertinent cases, bounds derived from the cLFV transitions are found to be more stringent in comparison to the low-energy observables, even in cases for which the relevant cLFV operator is generated using RGEs, as in the case of $\mu^\pm\to e^\pm e^+ e^- $ transition.

In conclusion, several promising directions come to light regarding future work. Moving away from the exact flavor symmetries presented here, an equally comprehensive analysis, which would instead be based on modular flavor symmetries~\cite{Kobayashi:2021pav}, may be performed. In conjunction with the lepton sector, yet another avenue for enhancing the present analysis could be the inclusion of the quark sector, either portrayed by the continuous or (modular) discrete flavor symmetries, which would broaden the list of the NP mediators considered in this letter.

Lastly, it has to be emphasized that the list of the well-motivated discrete flavor symmetries presented in this work is far from exhaustive. A discussion of a range of other viable discrete symmetry choices utilized in the model building is presented in Ref.~\cite{Ishimori:2010au}, which would motivate the future studies in an analogous context.


\begin{acknowledgments}
The author thanks Admir Greljo and Xavier Ponce D\'iaz for useful discussions and comments on the manuscript. The work received funding from the Swiss National Science Foundation (SNF) through the Eccellenza Professorial Fellowship ``Flavor Physics at the High Energy Frontier'' project number 186866.

\end{acknowledgments}


\begin{table*}[t]
\centering
\scalebox{0.80}{
\begin{tabular}{ccccc}
\toprule
\multicolumn{5}{c}{\textbf{$A_4$ Flavor symmetry}}
\\
\midrule
\textbf{UV Field}&\textbf{$-\cL_{\sscript{UV}}^{(4)}\supset$} & \textbf{Irrep}& \textbf{Invariants} &\textbf{Flavor tensor}
\\
\midrule
$\cS_1\sim(\bm1,\bm1)_{1}$
&$[y_{\mathcal S_1}]_{rij}\mathcal S_{1r}^\dag \bar\ell_{i}i\sig_2\ell_{j}^c$
&$\bm3$
&$[\bar\ell\,\ell^c]_{\bm{3_A}}[\cS_1]_{\bm3}$
&$[y_{\cS_1}]_{rij}=\frac{y_{\cS_1}}{2}\Big[ \delta_{r1}(\delta_{i2}\delta_{j3}-\delta_{i3}\delta_{j2})+\text{perm.} \Big]$
\\
\midrule
\multirow{3}{*}{\vspace{-0.25cm}$\cS_2\sim(\bm1,\bm1)_{2}$}
&\multirow{3}{*}{\vspace{-0.25cm}$[y_{\mathcal S_2}]_{rij}\mathcal S_{2r}^\dag \bar e_{i}e_{j}^c$}
&$\bm1$
&$[\bar e\,e^c]_{\bm{1}}[\cS_2]_{\bm1}$
&\vspace{+0.1cm}$[y_{\cS_2}]_{ij}=y_{\cS_2}^{(1)} \delta_{i1}\delta_{j1}+y_{\cS_2}^{(2)}(\delta_{i2}\delta_{j3}+\delta_{i3}\delta_{j2})$
\\
&
&$~\bm1'$
&$[\bar e\,e^c]_{\bm{1''}}[\cS_2]_{\bm1'}$
&\vspace{+0.1cm}$[y_{\cS_2}]_{ij}= y_{\cS_2}^{(1)}\delta_{i2}\delta_{j2}+y_{\cS_2}^{(2)}(\delta_{i1}\delta_{j3}+\delta_{i3}\delta_{j1})$
\\
&
&$~\,\bm1''$
&$[\bar e\,e^c]_{\bm{1'}}[\cS_2]_{\bm1''}$
&$[y_{\cS_2}]_{ij}= y_{\cS_2}^{(1)}\delta_{i3}\delta_{j3}+y_{\cS_2}^{(2)}(\delta_{i1}\delta_{j2}+\delta_{i2}\delta_{j1})$
\\
\midrule
\multirow{3}{*}{\vspace{-0.4cm}$\varphi\sim(\bm1,\bm2)_{\frac{1}{2}}$}
&\multirow{3}{*}{\vspace{-0.4cm}$[y_\varphi]_{rij}\varphi_r\bar\ell_ie_j$}
&\multirow{3}{*}{\vspace{-0.4cm}$\bm3$}
&\multirow{3}{*}{\vspace{-0.4cm}$[\bar\ell\,\varphi]_{\bm1}[e]_{\bm1}\oplus[\bar\ell\,\varphi]_{\bm1'}[e]_{\bm1''}\oplus[\bar\ell\,\varphi]_{\bm1''}[e]_{\bm1'}$}
&$[y_\varphi]_{rij}=\delta_{r1}\lzm y_\varphi^{(1)} \delta_{i1}\delta_{j1}+y_\varphi^{(2)}\delta_{i2}\delta_{j3}+y_\varphi^{(3)}\delta_{i3}\delta_{j2} \dzm$
\\
&&&&$~~~~~~~~~~+\delta_{r2}\lzm y_\varphi^{(1)} \delta_{i3}\delta_{j1}+y_\varphi^{(2)}\delta_{i1}\delta_{j3}+y_\varphi^{(3)}\delta_{i2}\delta_{j2} \dzm$
\\
&&&&$~~~~~~~~~~+\delta_{r3}\lzm y_\varphi^{(1)} \delta_{i2}\delta_{j1}+y_\varphi^{(2)}\delta_{i3}\delta_{j3}+y_\varphi^{(3)}\delta_{i1}\delta_{j2} \dzm$
\\
\midrule
\multirow{4}{*}{\vspace{-0.15cm}$\Xi_1\sim(\bm1,\bm3)_{1}$}
&\multirow{4}{*}{\vspace{-0.15cm}$[y_{\Xi_1}]_{rij}\Xi_{1r}^{a\dag}\bar\ell_{i}\sig^ai\sig_2\ell_{j}^c$}
&$\bm1$
&$[\bar\ell\,\ell^c]_{\bm1}[\Xi_1]_{\bm1}$
&$[y_{\Xi_1}]_{ij}=y_{\Xi_1}(\delta_{i1}\delta_{j1}+\delta_{i2}\delta_{j3}+\delta_{i3}\delta_{j2})$
\\
&
&$~\bm1'$
&$[\bar\ell\,\ell^c]_{\bm1''}[\Xi_1]_{\bm1'}$
&$[y_{\Xi_1}]_{ij}=y_{\Xi_1}(\delta_{i2}\delta_{j2}+\delta_{i1}\delta_{j3}+\delta_{i3}\delta_{j1})$
\\
&
&$~\,\bm1''$
&$[\bar\ell\,\ell^c]_{\bm1'}[\Xi_1]_{\bm1''}$
&$[y_{\Xi_1}]_{ij}=y_{\Xi_1}(\delta_{i3}\delta_{j3}+\delta_{i1}\delta_{j2}+\delta_{i2}\delta_{j1})$
\\
&
&$\bm3$
&$[\bar\ell\,\ell^c]_{\bm{3_S}}[\Xi_1]_{\bm3}$
&$[y_{\Xi_1}]_{rij}=\frac{y_{\Xi_1}}{3}\Big[\delta_{r1}(2\delta_{i1}\delta_{j1}-\delta_{i2}\delta_{j3}-\delta_{i3}\delta_{j2})+\text{perm.}\Big]$
\\
\midrule
\midrule
$N\sim(\bm1,\bm1)_0$
&$[\lambda_N]_{ri}\bar N_{R,r}\tilde\phi^\dag \ell_i$
&$\bm3$
&$[\bar N_R\,\ell]_{\bm1}$
&$[\lambda_N]_{ri}=\lambda_N(\delta_{r1}\delta_{i1}+\delta_{r2}\delta_{i3}+\delta_{r3}\delta_{i2})$
\\
\midrule
$E\sim(\bm1,\bm1)_{-1}$
&$[\lambda_E]_{ri}\bar E_{R,r}\phi^\dag\ell_{i}$
&$\bm3$
&$[\bar E_R\,\ell]_{\bm1}$
&$[\lambda_E]_{ri}=\lambda_E(\delta_{r1}\delta_{i1}+\delta_{r2}\delta_{i3}+\delta_{r3}\delta_{i2})$
\\
\midrule
\multirow{3}{*}{\vspace{-0.1cm}$\Delta_1\sim(\bm1,\bm2)_{-\frac{1}{2}}$}
&\multirow{3}{*}{\vspace{-0.1cm}$[\lambda_{\Delta_1}]_{ri}\bar\Delta_{1L,r}\phi e_i$}
&$\bm1$
&$[\Bar\Delta_{1L}]_{\bm1}[e]_{\bm1}$
&$[\lambda_{\Delta_1}]_i=\lambda_{\Delta_1}\delta_{i1}$
\\
&
&$~\bm1'$
&$[\Bar\Delta_{1L}]_{\bm1'}[e]_{\bm1''}$
&$[\lambda_{\Delta_1}]_i=\lambda_{\Delta_1}\delta_{i3}$
\\
&
&$~\,\bm1''$
&$[\Bar\Delta_{1L}]_{\bm1''}[e]_{\bm1'}$
&$[\lambda_{\Delta_1}]_i=\lambda_{\Delta_1}\delta_{i2}$
\\
\midrule
\multirow{3}{*}{\vspace{-0.1cm}$\Delta_3\sim(\bm1,\bm2)_{-\frac{3}{2}}$}
&\multirow{3}{*}{\vspace{-0.1cm}$[\lambda_{\Delta_3}]_{ri}\bar\Delta_{3L,r}\tilde\phi e_{i}$}
&$\bm1$
&$[\Bar\Delta_{3L}]_{\bm1}[e]_{\bm1}$
&$[\lambda_{\Delta_3}]_i=\lambda_{\Delta_3}\delta_{i1}$
\\
&
&$~\bm1'$
&$[\Bar\Delta_{3L}]_{\bm1'}[e]_{\bm1''}$
&$[\lambda_{\Delta_3}]_i=\lambda_{\Delta_3}\delta_{i3}$
\\
&
&$~\,\bm1''$
&$[\Bar\Delta_{3L}]_{\bm1''}[e]_{\bm1'}$
&$[\lambda_{\Delta_3}]_i=\lambda_{\Delta_3}\delta_{i2}$
\\
\midrule
$\Sigma\sim(\bm1,\bm3)_{0}$
&$\frac{1}{2}[\lambda_\Sigma]_{ri}\bar\Sigma^a_{R,r}\tilde\phi^\dag\sig^a\ell_{i}$
&$\bm3$
&$[\bar\Sigma_R\,\ell]_{\bm1}$
&$[\lambda_\Sigma]_{ri}=\lambda_\Sigma(\delta_{r1}\delta_{i1}+\delta_{r2}\delta_{i3}+\delta_{r3}\delta_{i2})$
\\
\midrule
$\Sigma_1\sim(\bm1,\bm3)_{-1}$
&$\frac{1}{2}[\lambda_{\Sigma_1}]_{ri}\bar\Sigma^a_{1R,r}\phi^\dag\sig^a\ell_{i}$
&$\bm3$
&$[\bar\Sigma_{1R}\,\ell]_{\bm1}$
&$[\lambda_{\Sigma_1}]_{ri}=\lambda_{\Sigma_1}(\delta_{r1}\delta_{i1}+\delta_{r2}\delta_{i3}+\delta_{r3}\delta_{i2})$
\\
\midrule
\midrule
\multirow{8}{*}{\vspace{-2.9cm}$\cB\sim(\bm1,\bm1)_0$}
&\multirow{8}{*}{\vspace{-2.3cm}$[g_\cB^\ell]_{rij}\cB_r^\mu\bar\ell_{i}\gamma_\mu\ell_{j}$}
&\multirow{2}{*}{$\bm1$}
&$[\bar\ell\,\ell]_{\bm1}[\cB]_{\bm1}$
&$[g_\cB^\ell]_{ij}=g_\cB^\ell(\delta_{i1}\delta_{j1}+\delta_{i2}\delta_{j2}+\delta_{i3}\delta_{j3})$
\\
&\multirow{8}{*}{\vspace{-2.6cm}$+[g_\cB^e]_{rij}\cB_r^\mu\bar e_{i}\gamma_\mu e_{j}~~$}
&
&$[\bar e\, e]_{\bm1}[\cB]_{\bm1}$
&$[g_\cB^e]_{ij}=g_\cB^{e(1)}\delta_{i1}\delta_{j1}+g_\cB^{e(2)}\delta_{i2}\delta_{j2}+g_\cB^{e(3)}\delta_{i3}\delta_{j3}$
\\
\cmidrule{3-5}
&&\multirow{2}{*}{$~\bm1'$}
&$[\bar\ell\,\ell]_{\bm1''}[\cB]_{\bm1'}$
&$[g_\cB^\ell]_{ij}=g_\cB^\ell(\delta_{i2}\delta_{j3}+\delta_{i1}\delta_{j2}+\delta_{i3}\delta_{j1})$
\\
&&&$[\bar e\, e]_{\bm1''}[\cB]_{\bm1'}$
&$[g_\cB^e]_{ij}=g_\cB^{e(1)}\delta_{i1}\delta_{j3}+g_\cB^{e(2)}\delta_{i2}\delta_{j1}+g_\cB^{e(3)}\delta_{i3}\delta_{j2}$
\\
\cmidrule{3-5}
&&\multirow{2}{*}{$~\,\bm1''$}
&$[\bar\ell\,\ell]_{\bm1'}[\cB]_{\bm1''}$
&$[g_\cB^\ell]_{ij}=g_\cB^\ell(\delta_{i3}\delta_{j2}+\delta_{i1}\delta_{j3}+\delta_{i2}\delta_{j1})$
\\
&&&$[\bar e\, e]_{\bm1'}[\cB]_{\bm1''}$
&$[g_\cB^e]_{ij}=g_\cB^{e(1)}\delta_{i1}\delta_{j2}+g_\cB^{e(2)}\delta_{i2}\delta_{j3}+g_\cB^{e(3)}\delta_{i3}\delta_{j1}$
\\
\cmidrule{3-5}
&&\multirow{3}{*}{\vspace{-0.1cm}$~\bm3_1$}
&\multirow{3}{*}{\vspace{-0.1cm}$[\bar\ell\,\ell]_{\bm{3_S}}[\cB]_{\bm3_1}$}
&$[g_\cB^\ell]_{rij}=\frac{g_\cB^\ell}{3}\Big[\delta_{r1}(2\delta_{i1}\delta_{j1}-\delta_{i2}\delta_{j2}-\delta_{i3}\delta_{j3})$
\\
&&&&$\,~~~~~~~~~~~~~~~+\delta_{r2}(2\delta_{i2}\delta_{j3}-\delta_{i1}\delta_{j2}-\delta_{i3}\delta_{j1})$
\\
&&&&$~~~~~~~~~~~~~~~~~+\delta_{r3}(2\delta_{i3}\delta_{j2}-\delta_{i1}\delta_{j3}-\delta_{i2}\delta_{j1})\Big]$
\\
\cmidrule{3-5}
&&\multirow{2}{*}{\vspace{-0.1cm}$~\bm3_2$}
&\multirow{2}{*}{\vspace{-0.1cm}$[\bar\ell\,\ell]_{\bm{3_A}}[\cB]_{\bm3_2}$}
&$[g_\cB^\ell]_{rij}=\frac{g_\cB^\ell}{2}\Big[\delta_{r1}(\delta_{i2}\delta_{j2}-\delta_{i3}\delta_{j3})+\delta_{r2}(\delta_{i1}\delta_{j3}-\delta_{i2}\delta_{j1})$
\\
&&&&$+\delta_{r3}(\delta_{i3}\delta_{j1}-\delta_{i1}\delta_{j2})\Big]~~~~~~~~~~~~$
\\
\midrule
\multirow{5}{*}{\vspace{-2.7cm}$\cW\sim(\bm1,\bm3)_0$}
&\multirow{5}{*}{\vspace{-2.7cm}$\frac{1}{2}[g_\cW^\ell]_{rij}\cW^{\mu\,a}_r\bar\ell_i\sigma^a\gamma_\mu\ell_j$}
&$\bm1$
&$[\bar\ell\,\ell]_{\bm1}[\cW]_{\bm1}$
&$[g_\cW^\ell]_{ij}=g_\cW^\ell(\delta_{i1}\delta_{j1}+\delta_{i2}\delta_{j2}+\delta_{i3}\delta_{j3})$
\\
&&$~\bm1'$
&$[\bar\ell\,\ell]_{\bm1''}[\cW]_{\bm1'}$
&$[g_\cW^\ell]_{ij}=g_\cW^\ell(\delta_{i2}\delta_{j3}+\delta_{i1}\delta_{j2}+\delta_{i3}\delta_{j1})$
\\
&&$~\,\bm1''$
&$[\bar\ell\,\ell]_{\bm1'}[\cW]_{\bm1''}$
&\vspace{+0.1cm}$[g_\cW^\ell]_{ij}=g_\cW^\ell(\delta_{i3}\delta_{j2}+\delta_{i1}\delta_{j3}+\delta_{i2}\delta_{j1})$
\\
\cmidrule{3-5}
&&\multirow{3}{*}{\vspace{-0.2cm}$~\bm3_1$}
&\multirow{3}{*}{\vspace{-0.2cm}$[\bar\ell\,\ell]_{\bm{3_S}}[\cW]_{\bm3_1}$}
&\vspace{+0.1cm}$[g_\cW^\ell]_{rij}=\frac{g_\cW^\ell}{3}\Big[\delta_{r1}(2\delta_{i1}\delta_{j1}-\delta_{i2}\delta_{j2}-\delta_{i3}\delta_{j3})$
\\
&&&&$\,~~~~~~~~~~~~~~~~+\delta_{r2}(2\delta_{i2}\delta_{j3}-\delta_{i1}\delta_{j2}-\delta_{i3}\delta_{j1})$
\\
&&&&$~~~~~~~~~~~~~~~~~~+\delta_{r3}(2\delta_{i3}\delta_{j2}-\delta_{i1}\delta_{j3}-\delta_{i2}\delta_{j1})\Big]$
\\
\cmidrule{3-5}
&&\multirow{2}{*}{\vspace{-0.1cm}$~\bm3_2$}
&\multirow{2}{*}{\vspace{-0.1cm}$[\bar\ell\,\ell]_{\bm{3_A}}[\cW]_{\bm3_1}$}
&$[g_\cW^\ell]_{rij}=\frac{g_\cW^\ell}{2}\Big[\delta_{r1}(\delta_{i2}\delta_{j2}-\delta_{i3}\delta_{j3})+\delta_{r2}(\delta_{i1}\delta_{j3}-\delta_{i2}\delta_{j1})$
\\
&&&&$+\delta_{r3}(\delta_{i3}\delta_{j1}-\delta_{i1}\delta_{j2})\Big]~~~~~~~~~~$
\\
\midrule
\multirow{3}{*}{\vspace{-0.6cm}$\cL_3\sim(\bm1,\bm2)_{-\frac{3}{2}}$}
&\multirow{3}{*}{\vspace{-0.6cm}$[g_{\cL_3}]_{rij}\cL_{3r}^{\mu\dag}\bar e_{i}^c\gamma_\mu \ell_{j}$}
&\multirow{3}{*}{\vspace{-0.6cm}$\bm3$}
&\multirow{3}{*}{\vspace{-0.6cm}$[\bar e^c]_{\bm1}[\ell\,\cL_3]_{\bm1}\oplus[\bar e^c]_{\bm1'}[\ell\,\cL_3]_{\bm1''}\oplus[\bar e^c]_{\bm1''}[\ell\,\cL_3]_{\bm1'}$}
&$[g_{\cL_3}]_{rij}=\delta_{r1}\lzm g_{\cL_3}^{(1)}\delta_{i1}\delta_{j1}+g_{\cL_3}^{(2)}\delta_{i2}\delta_{j3}+g_{\cL_3}^{(3)}\delta_{i3}\delta_{j2} \dzm$
\\
&&&
&$~~~~~~~~~~~+\delta_{r2}\lzm g_{\cL_3}^{(1)}\delta_{i3}\delta_{j1}+g_{\cL_3}^{(2)}\delta_{i1}\delta_{j3}+g_{\cL_3}^{(3)}\delta_{i2}\delta_{j2} \dzm$
\\
&&&
&$~~~~~~~~~~~+\delta_{r3}\lzm g_{\cL_3}^{(1)}\delta_{i2}\delta_{j1}+g_{\cL_3}^{(2)}\delta_{i3}\delta_{j3}+g_{\cL_3}^{(3)}\delta_{i1}\delta_{j2} \dzm$
\\[5pt]
\bottomrule
\end{tabular}
}
\caption{Classification of UV mediators under the $A_4$ flavor symmetry assumption. The first column lists the NP mediators along with their gauge quantum numbers. In the second column we indicate
the corresponding UV Lagrangian terms describing the interactions with the SM leptons. In the third and fourth column we list the possible irreps under the $A_4$ flavor symmetry and the corresponding flavor invariants following the notation introduced in Sec.~\ref{sec:discrete_flavor_symmetries}. Last column contains flavor tensors for a given flavor irrep. For $\cB$ and $\cW$ mediators, the symmetrization in the flavor tensors is implicitly assumed where appropriate.} \label{tab:A4 invariants and tensors}
\end{table*}

\begin{table*}[t]
\centering
\scalebox{0.95}{
\begin{tabular}{ccccc}
\toprule
\multicolumn{5}{c}{\textbf{$A_5$ Flavor symmetry}}
\\
\midrule
\textbf{UV Field}&\textbf{$-\cL_{\sscript{UV}}^{(4)}\supset$} & \textbf{Irrep}& \textbf{Invariants} &\textbf{Flavor tensor}
\\
\midrule
$\cS_1\sim(\bm1,\bm1)_{1}$
&$[y_{\mathcal S_1}]_{rij}\mathcal S_{1r}^\dag \bar\ell_{i}i\sig_2\ell_{j}^c$
&$\bm3$
&$[\bar\ell\,\ell^c]_{\bm{3_A}}[\cS_1]_{\bm3}$
&$[y_{\cS_1}]_{rij}=\frac{y_{\cS_1}}{2}\Big[ \delta_{r1}(\delta_{i2}\delta_{j3}-\delta_{i3}\delta_{j2})+\text{perm.} \Big]$
\\
\midrule
\multirow{3}{*}{\vspace{-0.3cm}$\cS_2\sim(\bm1,\bm1)_{2}$}
&\multirow{3}{*}{\vspace{-0.3cm}$[y_{\mathcal S_2}]_{rij}\mathcal S_{2r}^\dag \bar e_{i}e_{j}^c$}
&\multirow{3}{*}{\vspace{-0.3cm}$\bm1$}
&\multirow{3}{*}{\vspace{-0.3cm}$[\bar e\,e^c]_{\bm{1}}[\cS_2]_{\bm1}$}
&\vspace{+0.1cm}$~~~[y_{\cS_2}]_{ij}=y_{\cS_2}^{(1)}\delta_{i1}\delta_{j1}+y_{\cS_2}^{(2)}\delta_{i2}\delta_{j2}+y_{\cS_2}^{(3)}\delta_{i3}\delta_{j3}$
\\
&
&
&
&\vspace{+0.1cm}$~~~~~~~~~~~~+y_{\cS_2}^{(4)}\delta_{i1}\delta_{j2}+y_{\cS_2}^{(4)}\delta_{i2}\delta_{j1}+y_{\cS_2}^{(5)}\delta_{i1}\delta_{j3}$
\\
&
&
&
&$~~~~~~~~~~~~+y_{\cS_2}^{(5)}\delta_{i3}\delta_{j1}+y_{\cS_2}^{(6)}\delta_{i2}\delta_{j3}+y_{\cS_2}^{(6)}\delta_{i3}\delta_{j2}$
\\
\midrule
\multirow{3}{*}{\vspace{-0.4cm}$\varphi\sim(\bm1,\bm2)_{\frac{1}{2}}$}
&\multirow{3}{*}{\vspace{-0.4cm}$[y_\varphi]_{rij}\varphi_r\bar\ell_ie_j$}
&\multirow{3}{*}{\vspace{-0.4cm}$\bm3$}
&\multirow{3}{*}{\vspace{-0.4cm}$[\bar\ell\,\varphi]_{\bm1}[e]_{\bm1}$}
&$[y_\varphi]_{rij}=\delta_{r1}\lzm y_\varphi^{(1)} \delta_{i1}\delta_{j1}+y_\varphi^{(2)}\delta_{i1}\delta_{j2}+y_\varphi^{(3)}\delta_{i1}\delta_{j3} \dzm$
\\
&&&&$~~~~~~~~~~+\delta_{r2}\lzm y_\varphi^{(1)} \delta_{i3}\delta_{j1}+y_\varphi^{(2)}\delta_{i3}\delta_{j2}+y_\varphi^{(3)}\delta_{i3}\delta_{j3} \dzm$
\\
&&&&$~~~~~~~~~~+\delta_{r3}\lzm y_\varphi^{(1)} \delta_{i2}\delta_{j1}+y_\varphi^{(2)}\delta_{i2}\delta_{j2}+y_\varphi^{(3)}\delta_{i2}\delta_{j3} \dzm$
\\
\midrule
\multirow{4}{*}{\vspace{-0.6cm}$\Xi_1\sim(\bm1,\bm3)_{1}$}
&\multirow{4}{*}{\vspace{-0.6cm}$[y_{\Xi_1}]_{rij}\Xi_{1r}^{a\dag}\bar\ell_{i}\sig^ai\sig_2\ell_{j}^c$}
&$\bm1$
&$[\bar\ell\,\ell^c]_{\bm1}[\Xi_1]_{\bm1}$
&\vspace{+0.05cm}$[y_{\Xi_1}]_{ij}=y_{\Xi_1}(\delta_{i1}\delta_{j1}+\delta_{i2}\delta_{j3}+\delta_{i3}\delta_{j2})$
\\
\cmidrule{3-5}
&
&\multirow{3}{*}{\vspace{-0.3cm}$\bm5$}
&\multirow{3}{*}{\vspace{-0.3cm}$[\bar\ell\,\ell^c]_{\bm{5_S}}[\Xi_1]_{\bm5}$}
&\vspace{+0.05cm}$[y_{\Xi_1}]_{rij}=y_{\Xi_1}\Big[\delta_{r1}(2\delta_{i1}\delta_{j1}-\delta_{i2}\delta_{j3}-\delta_{i3}\delta_{j2})~~$
\\
&
&
&
&\vspace{+0.05cm}$~~~~~~~~~~~~~~~~~-\sqrt3\delta_{r2}(\delta_{i1}\delta_{j3}+\delta_{i3}\delta_{j1})+\sqrt6\delta_{r3}\delta_{i3}\delta_{j3}$
\\
&
&
&
&$~~~~~~~~~~~~~~~~~~+\sqrt6\delta_{r4}\delta_{i2}\delta_{j2}-\sqrt3\delta_{r5}(\delta_{i1}\delta_{j2}+\delta_{i2}\delta_{j1})\Big]$
\\
\midrule
\midrule
$N\sim(\bm1,\bm1)_0$
&$[\lambda_N]_{ri}\bar N_{R,r}\tilde\phi^\dag \ell_i$
&$\bm3$
&$[\bar N_R\,\ell]_{\bm1}$
&$[\lambda_N]_{ri}=\lambda_N(\delta_{r1}\delta_{i1}+\delta_{r2}\delta_{i3}+\delta_{r3}\delta_{i2})$
\\
\midrule
$E\sim(\bm1,\bm1)_{-1}$
&$[\lambda_E]_{ri}\bar E_{R,r}\phi^\dag\ell_{i}$
&$\bm3$
&$[\bar E_R\,\ell]_{\bm1}$
&$[\lambda_E]_{ri}=\lambda_E(\delta_{r1}\delta_{i1}+\delta_{r2}\delta_{i3}+\delta_{r3}\delta_{i2})$
\\
\midrule
\multirow{1}{*}{\vspace{-0.0cm}$\Delta_1\sim(\bm1,\bm2)_{-\frac{1}{2}}$}
&\multirow{1}{*}{\vspace{-0.0cm}$[\lambda_{\Delta_1}]_{ri}\bar\Delta_{1L,r}\phi e_i$}
&$\bm1$
&$[\Bar\Delta_{1L}]_{\bm1}[e]_{\bm1}$
&$[\lambda_{\Delta_1}]_i=\lambda_{\Delta_1}^{(1)}\delta_{i1}+\lambda_{\Delta_1}^{(2)}\delta_{i2}+\lambda_{\Delta_1}^{(3)}\delta_{i3}$
\\[3pt]
\midrule
\multirow{1}{*}{\vspace{-0.0cm}$\Delta_3\sim(\bm1,\bm2)_{-\frac{3}{2}}$}
&\multirow{1}{*}{\vspace{-0.0cm}$[\lambda_{\Delta_3}]_{ri}\bar\Delta_{3L,r}\tilde\phi e_{i}$}
&$\bm1$
&$[\Bar\Delta_{3L}]_{\bm1}[e]_{\bm1}$
&$[\lambda_{\Delta_3}]_i=\lambda_{\Delta_3}^{(1)}\delta_{i1}+\lambda_{\Delta_3}^{(2)}\delta_{i2}+\lambda_{\Delta_3}^{(3)}\delta_{i3}$
\\[3pt]
\midrule
$\Sigma\sim(\bm1,\bm3)_{0}$
&$\frac{1}{2}[\lambda_\Sigma]_{ri}\bar\Sigma^a_{R,r}\tilde\phi^\dag\sig^a\ell_{i}$
&$\bm3$
&$[\bar\Sigma_R\,\ell]_{\bm1}$
&$[\lambda_\Sigma]_{ri}=\lambda_\Sigma(\delta_{r1}\delta_{i1}+\delta_{r2}\delta_{i3}+\delta_{r3}\delta_{i2})$
\\
\midrule
$\Sigma_1\sim(\bm1,\bm3)_{-1}$
&$\frac{1}{2}[\lambda_{\Sigma_1}]_{ri}\bar\Sigma^a_{1R,r}\phi^\dag\sig^a\ell_{i}$
&$\bm3$
&$[\bar\Sigma_{1R}\,\ell]_{\bm1}$
&$[\lambda_{\Sigma_1}]_{ri}=\lambda_{\Sigma_1}(\delta_{r1}\delta_{i1}+\delta_{r2}\delta_{i3}+\delta_{r3}\delta_{i2})$
\\
\midrule
\midrule
\multirow{8}{*}{\vspace{-1.45cm}$\cB\sim(\bm1,\bm1)_0$}
&\multirow{8}{*}{\vspace{-0.9cm}$[g_\cB^\ell]_{rij}\cB_r^\mu\bar\ell_{i}\gamma_\mu\ell_{j}$}
&\multirow{3}{*}{\vspace{-0.5cm}$\bm1$}
&\vspace{+0.05cm}$[\bar\ell\,\ell]_{\bm1}[\cB]_{\bm1}$
&$[g_\cB^\ell]_{ij}=g_\cB^\ell(\delta_{i1}\delta_{j1}+\delta_{i2}\delta_{j2}+\delta_{i3}\delta_{j3})$
\\
\cmidrule{4-5}
&\multirow{8}{*}{\vspace{-0.6cm}$+[g_\cB^e]_{rij}\cB_r^\mu\bar e_{i}\gamma_\mu e_{j}~~$}
&
&\multirow{2}{*}{$[\bar e\, e]_{\bm1}[\cB]_{\bm1}$}
&$[g_\cB^e]_{ij}=g_\cB^{e(1)}\delta_{i1}\delta_{j1}+g_\cB^{e(2)}\delta_{i2}\delta_{j2}+g_\cB^{e(3)}\delta_{i3}\delta_{j3}$
\\
&&&&$~~~~~~~~~+g_\cB^{e(4)}\delta_{i1}\delta_{j2}+g_\cB^{e(5)}\delta_{i1}\delta_{j3}+g_\cB^{e(6)}\delta_{i2}\delta_{j3}$
\\
\cmidrule{3-5}
&&\multirow{2}{*}{\vspace{-0.2cm}$\bm3$}
&\multirow{2}{*}{\vspace{-0.2cm}$[\bar\ell\,\ell]_{\bm{3_A}}[\cB]_{\bm3}$}
&$[g_\cB^\ell]_{rij}=g_\cB^\ell\Big[\delta_{r1}(\delta_{i2}\delta_{j2}-\delta_{i3}\delta_{j3})+\delta_{r2}(\delta_{i3}\delta_{j1}-\delta_{i1}\delta_{j2})$
\\
&&&
&$+\delta_{r3}(\delta_{i1}\delta_{j3}-\delta_{i2}\delta_{j1})\Big]~~~~~~~~~~~~~$
\\
\cmidrule{3-5}
&&\multirow{3}{*}{\vspace{-0.2cm}$\bm5$}
&\multirow{3}{*}{\vspace{-0.2cm}$[\bar\ell\,\ell]_{\bm{5_S}}[\cB]_{\bm5}$}
&$[g_\cB^\ell]_{rij}=g_\cB^\ell\Big[\delta_{r1}(2\delta_{i1}\delta_{j1}-\delta_{i2}\delta_{j2}-\delta_{i3}\delta_{j3})~~~~~~$
\\
&&&
&$~~~~~~~~~~-\sqrt3\delta_{r2}(\delta_{i1}\delta_{j2}+\delta_{i3}\delta_{j1})+\sqrt6\delta_{r3}\delta_{i3}\delta_{j2}$
\\
&&&
&$~~~~~~~~~~~+\sqrt6\delta_{r4}\delta_{i2}\delta_{j3}-\sqrt3\delta_{r5}(\delta_{i1}\delta_{j3}+\delta_{i2}\delta_{j1})\Big]$
\\
\midrule
\multirow{5}{*}{\vspace{-1.7cm}$\cW\sim(\bm1,\bm3)_0$}
&\multirow{5}{*}{\vspace{-1.7cm}$\frac{1}{2}[g_\cW^\ell]_{rij}\cW^{\mu\,a}_r\bar\ell_i\sigma^a\gamma_\mu\ell_j$}
&$\bm1$
&$[\bar\ell\,\ell]_{\bm1}[\cW]_{\bm1}$
&$[g_\cW^\ell]_{ij}=g_\cW^\ell(\delta_{i1}\delta_{j1}+\delta_{i2}\delta_{j2}+\delta_{i3}\delta_{j3})$
\\
\cmidrule{3-5}
&&\multirow{2}{*}{\vspace{-0.2cm}$\bm3$}
&\multirow{2}{*}{\vspace{-0.2cm}$[\bar\ell\,\ell]_{\bm{3_A}}[\cW]_{\bm3}$}
&$[g_\cW^\ell]_{rij}=g_\cW^\ell\Big[\delta_{r1}(\delta_{i2}\delta_{j2}-\delta_{i3}\delta_{j3})+\delta_{r2}(\delta_{i3}\delta_{j1}-\delta_{i1}\delta_{j2})$
\\
&&&
&$+\delta_{r3}(\delta_{i1}\delta_{j3}-\delta_{i2}\delta_{j1})\Big]~~~~~~~~~~~~~$
\\
\cmidrule{3-5}
&&\multirow{3}{*}{$\bm5$}
&\multirow{3}{*}{$[\bar\ell\,\ell]_{\bm{5_S}}[\cW]_{\bm5}$}
&$[g_\cW^\ell]_{rij}=g_\cW^\ell\Big[\delta_{r1}(2\delta_{i1}\delta_{j1}-\delta_{i2}\delta_{j2}-\delta_{i3}\delta_{j3})~~~~~~$
\\
&&&&$~~~~~~~~~~-\sqrt3\delta_{r2}(\delta_{i1}\delta_{j2}+\delta_{i3}\delta_{j1})+\sqrt6\delta_{r3}\delta_{i3}\delta_{j2}$
\\
&&&&$~~~~~~~~~~~+\sqrt6\delta_{r4}\delta_{i2}\delta_{j3}-\sqrt3\delta_{r5}(\delta_{i1}\delta_{j3}+\delta_{i2}\delta_{j1})\Big]$
\\
\midrule
\multirow{3}{*}{\vspace{-0.6cm}$\cL_3\sim(\bm1,\bm2)_{-\frac{3}{2}}$}
&\multirow{3}{*}{\vspace{-0.6cm}$[g_{\cL_3}]_{rij}\cL_{3r}^{\mu\dag}\bar e_{i}^c\gamma_\mu \ell_{j}$}
&\multirow{3}{*}{\vspace{-0.6cm}$\bm3$}
&\multirow{3}{*}{\vspace{-0.6cm}$[\bar e^c]_{\bm1}[\ell\,\cL_3]_{\bm1}$}
&$[g_{\cL_3}]_{rij}=\delta_{r1}\lzm g_{\cL_3}^{(1)}\delta_{i1}\delta_{j1}+g_{\cL_3}^{(2)}\delta_{i2}\delta_{j1}+g_{\cL_3}^{(3)}\delta_{i3}\delta_{j1} \dzm$
\\
&&&
&$~~~~~~~~~~~+\delta_{r2}\lzm g_{\cL_3}^{(1)}\delta_{i1}\delta_{j3}+g_{\cL_3}^{(2)}\delta_{i2}\delta_{j3}+g_{\cL_3}^{(3)}\delta_{i3}\delta_{j3} \dzm$
\\
&&&
&$~~~~~~~~~~~+\delta_{r3}\lzm g_{\cL_3}^{(1)}\delta_{i1}\delta_{j2}+g_{\cL_3}^{(2)}\delta_{i2}\delta_{j2}+g_{\cL_3}^{(3)}\delta_{i3}\delta_{j2} \dzm$
\\[5pt]
\bottomrule
\end{tabular}
}
\caption{Classification of UV mediators under the $A_5$ flavor symmetry assumption. The first column lists the NP mediators along with their gauge quantum numbers. In the second column we indicate
the corresponding UV Lagrangian terms describing the interactions with the SM leptons. In the third and fourth column we list the possible irreps under the $A_5$ flavor symmetry and the corresponding flavor invariants following the notation introduced in Sec.~\ref{sec:discrete_flavor_symmetries}. Last column contains flavor tensors for a given flavor irrep. For $\cB$ and $\cW$ mediators, the symmetrization in the flavor tensors is implicitly assumed where appropriate.}
\label{tab:A5 invariants and tensors}
\end{table*}

\begin{table*}[t]
\centering
\scalebox{0.95}{
\begin{tabular}{ccccc}
\toprule
\multicolumn{5}{c}{\textbf{$S_4$ Flavor symmetry}}
\\
\midrule
\textbf{UV Field}&\textbf{$-\cL_{\sscript{UV}}^{(4)}\supset$} & \textbf{Irrep}& \textbf{Invariants} &\textbf{Flavor tensor}
\\
\midrule
$\cS_1\sim(\bm1,\bm1)_{1}$
&$[y_{\mathcal S_1}]_{rij}\mathcal S_{1r}^\dag \bar\ell_{i}i\sig_2\ell_{j}^c$
&$\bm3$
&$[\bar\ell\,\ell^c]_{\bm{3_A}}[\cS_1]_{\bm3}$
&$[y_{\cS_1}]_{rij}=\frac{y_{\cS_1}}{2}\Big[ \delta_{r1}(\delta_{i2}\delta_{j3}-\delta_{i3}\delta_{j2})+\text{perm.} \Big]$
\\
\midrule
\multirow{3}{*}{\vspace{-0.3cm}$\cS_2\sim(\bm1,\bm1)_{2}$}
&\multirow{3}{*}{\vspace{-0.3cm}$[y_{\mathcal S_2}]_{rij}\mathcal S_{2r}^\dag \bar e_{i}e_{j}^c$}
&\multirow{3}{*}{\vspace{-0.3cm}$\bm1$}
&\multirow{3}{*}{\vspace{-0.3cm}$[\bar e\,e^c]_{\bm{1}}[\cS_2]_{\bm1}$}
&\vspace{+0.1cm}$~~~[y_{\cS_2}]_{ij}=y_{\cS_2}^{(1)}\delta_{i1}\delta_{j1}+y_{\cS_2}^{(2)}\delta_{i2}\delta_{j2}+y_{\cS_2}^{(3)}\delta_{i3}\delta_{j3}+y_{\cS_2}^{(4)}\delta_{i1}\delta_{j2}$
\\
&
&
&
&\vspace{+0.1cm}$~~~~~~~~~~~~~+y_{\cS_2}^{(4)}\delta_{i2}\delta_{j1}+y_{\cS_2}^{(5)}\delta_{i1}\delta_{j3}+y_{\cS_2}^{(5)}\delta_{i3}\delta_{j1}+y_{\cS_2}^{(6)}\delta_{i2}\delta_{j3}$
\\
&
&
&
&$~~+y_{\cS_2}^{(6)}\delta_{i3}\delta_{j2}~~~~~~~~~~~~~~~~~~~~~~~~~~~~~~~~~~~~$
\\
\midrule
\multirow{3}{*}{\vspace{-0.4cm}$\varphi\sim(\bm1,\bm2)_{\frac{1}{2}}$}
&\multirow{3}{*}{\vspace{-0.4cm}$[y_\varphi]_{rij}\varphi_r\bar\ell_ie_j$}
&\multirow{3}{*}{\vspace{-0.4cm}$\bm3$}
&\multirow{3}{*}{\vspace{-0.4cm}$[\bar\ell\,\varphi]_{\bm1}[e]_{\bm1}$}
&$[y_\varphi]_{rij}=\delta_{r1}\lzm y_\varphi^{(1)} \delta_{i1}\delta_{j1}+y_\varphi^{(2)}\delta_{i1}\delta_{j2}+y_\varphi^{(3)}\delta_{i1}\delta_{j3} \dzm$
\\
&&&&$~~~~~~~~~~+\delta_{r2}\lzm y_\varphi^{(1)} \delta_{i2}\delta_{j1}+y_\varphi^{(2)}\delta_{i2}\delta_{j2}+y_\varphi^{(3)}\delta_{i2}\delta_{j3} \dzm$
\\
&&&&$~~~~~~~~~~+\delta_{r3}\lzm y_\varphi^{(1)} \delta_{i3}\delta_{j1}+y_\varphi^{(2)}\delta_{i3}\delta_{j2}+y_\varphi^{(3)}\delta_{i3}\delta_{j3} \dzm$
\\
\midrule
\multirow{4}{*}{\vspace{-1.2cm}$\Xi_1\sim(\bm1,\bm3)_{1}$}
&\multirow{4}{*}{\vspace{-1.2cm}$[y_{\Xi_1}]_{rij}\Xi_{1r}^{a\dag}\bar\ell_{i}\sig^ai\sig_2\ell_{j}^c$}
&$\bm1$
&$[\bar\ell\,\ell^c]_{\bm1}[\Xi_1]_{\bm1}$
&\vspace{+0.05cm}$[y_{\Xi_1}]_{ij}=y_{\Xi_1}(\delta_{i1}\delta_{j1}+\delta_{i2}\delta_{j2}+\delta_{i3}\delta_{j3})$
\\
\cmidrule{3-5}
&
&\multirow{2}{*}{\vspace{-0.3cm}$\bm2$}
&\multirow{2}{*}{\vspace{-0.3cm}$[\bar\ell\,\ell^c]_{\bm{2}}[\Xi_1]_{\bm2}$}
&\vspace{+0.05cm}$[y_{\Xi_1}]_{rij}=y_{\Xi_1}\Big[\frac{1}{\sqrt2}\delta_{r1}(\delta_{i2}\delta_{j2}-\delta_{i3}\delta_{j3})~~~~~$
\\
&
&
&
&$~~~~~~~~~~~~~~~+\frac{1}{\sqrt6}\delta_{r2}(-2\delta_{i1}\delta_{j1}+\delta_{i2}\delta_{j2}+\delta_{i3}\delta_{j3})\Big]$
\\
\cmidrule{3-5}
&&$\bm3$
&$[\bar\ell\,\ell^c]_{\bm{3_S}}[\Xi_1]_{\bm3}$
&$[y_{\Xi_1}]_{rij}=y_{\Xi_1}\Big[\delta_{r1}(\delta_{i2}\delta_{j3}+\delta_{i3}\delta_{j2})+\text{perm.}\Big]$
\\[0.1cm]
\midrule
\midrule
$N\sim(\bm1,\bm1)_0$
&$[\lambda_N]_{ri}\bar N_{R,r}\tilde\phi^\dag \ell_i$
&$\bm3$
&$[\bar N_R\,\ell]_{\bm1}$
&$[\lambda_N]_{ri}=\lambda_N(\delta_{r1}\delta_{i1}+\delta_{r2}\delta_{i2}+\delta_{r3}\delta_{i3})$
\\
\midrule
$E\sim(\bm1,\bm1)_{-1}$
&$[\lambda_E]_{ri}\bar E_{R,r}\phi^\dag\ell_{i}$
&$\bm3$
&$[\bar E_R\,\ell]_{\bm1}$
&$[\lambda_E]_{ri}=\lambda_E(\delta_{r1}\delta_{i1}+\delta_{r2}\delta_{i2}+\delta_{r3}\delta_{i3})$
\\
\midrule
\multirow{1}{*}{\vspace{-0.0cm}$\Delta_1\sim(\bm1,\bm2)_{-\frac{1}{2}}$}
&\multirow{1}{*}{\vspace{-0.0cm}$[\lambda_{\Delta_1}]_{ri}\bar\Delta_{1L,r}\phi e_i$}
&$\bm1$
&$[\Bar\Delta_{1L}]_{\bm1}[e]_{\bm1}$
&$[\lambda_{\Delta_1}]_i=\lambda_{\Delta_1}^{(1)}\delta_{i1}+\lambda_{\Delta_1}^{(2)}\delta_{i2}+\lambda_{\Delta_1}^{(3)}\delta_{i3}$
\\[3pt]
\midrule
\multirow{1}{*}{\vspace{-0.0cm}$\Delta_3\sim(\bm1,\bm2)_{-\frac{3}{2}}$}
&\multirow{1}{*}{\vspace{-0.0cm}$[\lambda_{\Delta_3}]_{ri}\bar\Delta_{3L,r}\tilde\phi e_{i}$}
&$\bm1$
&$[\Bar\Delta_{3L}]_{\bm1}[e]_{\bm1}$
&$[\lambda_{\Delta_3}]_i=\lambda_{\Delta_3}^{(1)}\delta_{i1}+\lambda_{\Delta_3}^{(2)}\delta_{i2}+\lambda_{\Delta_3}^{(3)}\delta_{i3}$
\\[3pt]
\midrule
$\Sigma\sim(\bm1,\bm3)_{0}$
&$\frac{1}{2}[\lambda_\Sigma]_{ri}\bar\Sigma^a_{R,r}\tilde\phi^\dag\sig^a\ell_{i}$
&$\bm3$
&$[\bar\Sigma_R\,\ell]_{\bm1}$
&$[\lambda_\Sigma]_{ri}=\lambda_\Sigma(\delta_{r1}\delta_{i1}+\delta_{r2}\delta_{i2}+\delta_{r3}\delta_{i3})$
\\
\midrule
$\Sigma_1\sim(\bm1,\bm3)_{-1}$
&$\frac{1}{2}[\lambda_{\Sigma_1}]_{ri}\bar\Sigma^a_{1R,r}\phi^\dag\sig^a\ell_{i}$
&$\bm3$
&$[\bar\Sigma_{1R}\,\ell]_{\bm1}$
&$[\lambda_{\Sigma_1}]_{ri}=\lambda_{\Sigma_1}(\delta_{r1}\delta_{i1}+\delta_{r2}\delta_{i2}+\delta_{r3}\delta_{i3})$
\\
\midrule
\midrule
\multirow{8}{*}{\vspace{-0.2cm}$\cB\sim(\bm1,\bm1)_0$}
&\multirow{8}{*}{\vspace{+0.3cm}$[g_\cB^\ell]_{rij}\cB_r^\mu\bar\ell_{i}\gamma_\mu\ell_{j}$}
&\multirow{3}{*}{\vspace{-0.5cm}$\bm1$}
&\vspace{+0.05cm}$[\bar\ell\,\ell]_{\bm1}[\cB]_{\bm1}$
&$[g_\cB^\ell]_{ij}=g_\cB^\ell(\delta_{i1}\delta_{j1}+\delta_{i2}\delta_{j2}+\delta_{i3}\delta_{j3})$
\\
\cmidrule{4-5}
&\multirow{8}{*}{\vspace{+0.6cm}$+[g_\cB^e]_{rij}\cB_r^\mu\bar e_{i}\gamma_\mu e_{j}~~$}
&
&\multirow{2}{*}{$[\bar e\, e]_{\bm1}[\cB]_{\bm1}$}
&$[g_\cB^e]_{ij}=g_\cB^{e(1)}\delta_{i1}\delta_{j1}+g_\cB^{e(2)}\delta_{i2}\delta_{j2}+g_\cB^{e(3)}\delta_{i3}\delta_{j3}$
\\
&&&&$~~~~~~~~~+g_\cB^{e(4)}\delta_{i1}\delta_{j2}+g_\cB^{e(5)}\delta_{i1}\delta_{j3}+g_\cB^{e(6)}\delta_{i2}\delta_{j3}$
\\
\cmidrule{3-5}
&&\multirow{2}{*}{\vspace{-0.2cm}$\bm2$}
&\multirow{2}{*}{\vspace{-0.2cm}$[\bar\ell\,\ell]_{\bm{2}}[\cB]_{\bm2}$}
&$[g_\cB^\ell]_{rij}=g_\cB^\ell\Big[\frac{1}{\sqrt2}\delta_{r1}(\delta_{i2}\delta_{j2}-\delta_{i3}\delta_{j3})~~~~~$
\\
&&&
&$~~~~~~~~~~~~~~~+\frac{1}{\sqrt6}\delta_{r2}(-2\delta_{i1}\delta_{j1}+\delta_{i2}\delta_{j2}+\delta_{i3}\delta_{j3})\Big]$
\\
\cmidrule{3-5}
&&\multirow{1}{*}{\vspace{-0.0cm}$\bm3$}
&\multirow{1}{*}{\vspace{-0.0cm}$[\bar\ell\,\ell]_{\bm{3_S}}[\cB]_{\bm3}$}
&$[g_\cB^\ell]_{rij}=g_\cB^\ell\Big[\delta_{r1}(\delta_{i2}\delta_{j3}+\delta_{i3}\delta_{j2})+\text{perm.}\Big]$
\\[0.1cm]
\midrule
\multirow{5}{*}{\vspace{-0.4cm}$\cW\sim(\bm1,\bm3)_0$}
&\multirow{5}{*}{\vspace{-0.4cm}$\frac{1}{2}[g_\cW^\ell]_{rij}\cW^{\mu\,a}_r\bar\ell_i\sigma^a\gamma_\mu\ell_j$}
&$\bm1$
&$[\bar\ell\,\ell]_{\bm1}[\cW]_{\bm1}$
&$[g_\cW^\ell]_{ij}=g_\cW^\ell(\delta_{i1}\delta_{j1}+\delta_{i2}\delta_{j2}+\delta_{i3}\delta_{j3})$
\\
\cmidrule{3-5}
&&\multirow{2}{*}{\vspace{-0.2cm}$\bm2$}
&\multirow{2}{*}{\vspace{-0.2cm}$[\bar\ell\,\ell]_{\bm{2}}[\cW]_{\bm2}$}
&$[g_\cW^\ell]_{rij}=g_\cW^\ell\Big[\frac{1}{\sqrt2}\delta_{r1}(\delta_{i2}\delta_{j2}-\delta_{i3}\delta_{j3})~~~~~$
\\
&&&
&$~~~~~~~~~~~~~~~+\frac{1}{\sqrt6}\delta_{r2}(-2\delta_{i1}\delta_{j1}+\delta_{i2}\delta_{j2}+\delta_{i3}\delta_{j3})\Big]$
\\
\cmidrule{3-5}
&&\multirow{1}{*}{$\bm3$}
&\multirow{1}{*}{$[\bar\ell\,\ell]_{\bm{3_S}}[\cW]_{\bm3}$}
&$[g_\cW^\ell]_{rij}=g_\cW^\ell \Big[\delta_{r1}(\delta_{i2}\delta_{j3}+\delta_{i3}\delta_{j2})+\text{perm.}\Big]$
\\[0.1cm]
\midrule
\multirow{3}{*}{\vspace{-0.6cm}$\cL_3\sim(\bm1,\bm2)_{-\frac{3}{2}}$}
&\multirow{3}{*}{\vspace{-0.6cm}$[g_{\cL_3}]_{rij}\cL_{3r}^{\mu\dag}\bar e_{i}^c\gamma_\mu \ell_{j}$}
&\multirow{3}{*}{\vspace{-0.6cm}$\bm3$}
&\multirow{3}{*}{\vspace{-0.6cm}$[\bar e^c]_{\bm1}[\ell\,\cL_3]_{\bm1}$}
&$[g_{\cL_3}]_{rij}=\delta_{r1}\lzm g_{\cL_3}^{(1)}\delta_{i1}\delta_{j1}+g_{\cL_3}^{(2)}\delta_{i2}\delta_{j1}+g_{\cL_3}^{(3)}\delta_{i3}\delta_{j1} \dzm$
\\
&&&
&$~~~~~~~~~~~+\delta_{r2}\lzm g_{\cL_3}^{(1)}\delta_{i1}\delta_{j2}+g_{\cL_3}^{(2)}\delta_{i2}\delta_{j2}+g_{\cL_3}^{(3)}\delta_{i3}\delta_{j2} \dzm$
\\
&&&
&$~~~~~~~~~~~+\delta_{r3}\lzm g_{\cL_3}^{(1)}\delta_{i1}\delta_{j3}+g_{\cL_3}^{(2)}\delta_{i2}\delta_{j3}+g_{\cL_3}^{(3)}\delta_{i3}\delta_{j3} \dzm$
\\[5pt]
\bottomrule
\end{tabular}
}
\caption{Classification of UV mediators under the $S_4$ flavor symmetry assumption. The first column lists the NP mediators along with their gauge quantum numbers. In the second column we indicate
the corresponding UV Lagrangian terms describing the interactions with the SM leptons. In the third and fourth column we list the possible irreps under the $S_4$ flavor symmetry and the corresponding flavor invariants following the notation introduced in Sec.~\ref{sec:discrete_flavor_symmetries}. Last column contains flavor tensors for a given flavor irrep. For $\cB$ and $\cW$ mediators, the symmetrization in the flavor tensors is implicitly assumed where appropriate.}
\label{tab:S4 invariants and tensors}
\end{table*}

\begin{table*}[t]
\centering
\scalebox{0.86}{
\begin{tabular}{ccc}
\toprule
\multicolumn{3}{c}{\textbf{$A_4$ SMEFT Matching}}
\\
\midrule
\textbf{UV Field} & \textbf{Irrep}&\textbf{$\cL_{\sscript{SMEFT}}\supset$}
\\
\midrule
$\cS_1\sim(\bm1,\bm1)_{1}$
&$\bm3$
&$\frac{|y_{\cS_1}|^2}{2M^2_{\cS_1}}\Big[ [\cO_{\ell\ell}]_{1122}-[\cO_{\ell\ell}]_{1221}+[\cO_{\ell\ell}]_{1133}
			-[\cO_{\ell\ell}]_{1331}+[\cO_{\ell\ell}]_{2233}-[\cO_{\ell\ell}]_{2332} \Big]$
\\
\midrule
\multirow{5}{*}{\vspace{-2.1cm}$\Xi_1\sim(\bm1,\bm3)_{1}$}
&$\bm1$
&\vspace{+0.1cm}$\frac{|y_{\Xi_1}|^2}{M^2_{\Xi_1}}\Big[ [\cO_{\ell\ell}]_{1111}+2[\cO_{\ell\ell}]_{2233}+2[\cO_{\ell\ell}]_{2332} +(2[\cO_{\ell\ell}]_{1213}+\hermc)\Big]$
\\
&$~\bm1'$
&\vspace{+0.1cm}$\frac{|y_{\Xi_1}|^2}{M^2_{\Xi_1}}\Big[ [\cO_{\ell\ell}]_{2222}+2[\cO_{\ell\ell}]_{1133}+2[\cO_{\ell\ell}]_{1331} +(2[\cO_{\ell\ell}]_{1232}+\hermc)\Big]$
\\
&$~\,\bm1''$
&\vspace{+0.2cm}$\frac{|y_{\Xi_1}|^2}{M^2_{\Xi_1}}\Big[ [\cO_{\ell\ell}]_{3333}+2[\cO_{\ell\ell}]_{1122}+2[\cO_{\ell\ell}]_{1221} +(2[\cO_{\ell\ell}]_{1323}+\hermc)\Big]$
\\
\cmidrule{2-3}
&\multirow{2}{*}{\vspace{-0.5cm}$\,\bm3$}
&\vspace{+0.1cm}$\frac{4|y_{\Xi_1}|^2}{9M^2_{\Xi_1}}\Big[[\cO_{\ell\ell}]_{1111}+[\cO_{\ell\ell}]_{2222}+[\cO_{\ell\ell}]_{3333}\Big]+\frac{2|y_{\Xi_1}|^2}{9M^2_{\Xi_1}}\Big[[\cO_{\ell\ell}]_{1122}+[\cO_{\ell\ell}]_{1133}+[\cO_{\ell\ell}]_{2233}$
\\
&
&\vspace{+0.1cm}$+[\cO_{\ell\ell}]_{1221}+[\cO_{\ell\ell}]_{1331}+[\cO_{\ell\ell}]_{2332}\Big]-\frac{4|y_{\Xi_1}|^2}{9M^2_{\Xi_1}}\Big[[\cO_{\ell\ell}]_{1213}+[\cO_{\ell\ell}]_{1232}+[\cO_{\ell\ell}]_{1323}+\hermc\Big]$
\\
\midrule
\midrule
$N\sim(\bm1,\bm1)_0$
&$\bm3$
&\vspace{+0.1cm}$\frac{|\lambda_N|^2}{4M^2_N}\Big[[\cO_{\phi\ell}^{(1)}]_{11}-[\cO_{\phi\ell}^{(3)}]_{11}+[\cO_{\phi\ell}^{(1)}]_{22}-[\cO_{\phi\ell}^{(3)}]_{22}+[\cO_{\phi\ell}^{(1)}]_{33}-[\cO_{\phi\ell}^{(3)}]_{33}\Big]$
\\
\midrule
$E\sim(\bm1,\bm1)_{-1}$
&$\bm3$
&$-\frac{|\lambda_E|^2}{4M^2_E}\Big[[\cO_{\phi\ell}^{(1)}]_{11}+[\cO_{\phi\ell}^{(3)}]_{11}+[\cO_{\phi\ell}^{(1)}]_{22}+[\cO_{\phi\ell}^{(3)}]_{22}+[\cO_{\phi\ell}^{(1)}]_{33}+[\cO_{\phi\ell}^{(3)}]_{33}\Big]$
\\
\midrule
\multirow{1}{*}{\vspace{-0.0cm}$\Delta_1\sim(\bm1,\bm2)_{-\frac{1}{2}}$}
&$\bm1\oplus\bm1'\oplus\bm1''$
&\vspace{+0.1cm}$\lzs\frac{|\lambda_{\Delta_1}|^2}{2M^2_{\Delta_1}}[\cO_{\phi e}]_{11}\dzs_{\bm1}\oplus\lzs\frac{|\lambda_{\Delta_1}|^2}{2M^2_{\Delta_1}}[\cO_{\phi e}]_{33}\dzs_{\bm1'}\oplus\lzs\frac{|\lambda_{\Delta_1}|^2}{2M^2_{\Delta_1}}[\cO_{\phi e}]_{22}\dzs_{\bm1''}$
\\
\midrule
\multirow{1}{*}{\vspace{-0.0cm}$\Delta_3\sim(\bm1,\bm2)_{-\frac{3}{2}}$}
&$\bm1\oplus\bm1'\oplus\bm1''$
&\vspace{+0.1cm}$\lzs-\frac{|\lambda_{\Delta_3}|^2}{2M^2_{\Delta_3}}[\cO_{\phi e}]_{11}\dzs_{\bm1}\oplus\lzs-\frac{|\lambda_{\Delta_3}|^2}{2M^2_{\Delta_3}}[\cO_{\phi e}]_{33}\dzs_{\bm1'}\oplus\lzs-\frac{|\lambda_{\Delta_3}|^2}{2M^2_{\Delta_3}}[\cO_{\phi e}]_{22}\dzs_{\bm1''}$
\\
\midrule
$\Sigma\sim(\bm1,\bm3)_{0}$
&$\bm3$
&$\frac{|\lambda_\Sigma|^2}{16M^2_\Sigma}\Big[[\cO_{\phi\ell}^{(3)}]_{11}+3[\cO_{\phi\ell}^{(1)}]_{11}+[\cO_{\phi\ell}^{(3)}]_{22}+3[\cO_{\phi\ell}^{(1)}]_{22}+[\cO_{\phi\ell}^{(3)}]_{33}+3[\cO_{\phi\ell}^{(1)}]_{33}\Big]$
\\
\midrule
$\Sigma_1\sim(\bm1,\bm3)_{-1}$
&$\bm3$
&$\frac{|\lambda_{\Sigma_1}|^2}{16M^2_{\Sigma_1}}\Big[[\cO_{\phi\ell}^{(3)}]_{11}-3[\cO_{\phi\ell}^{(1)}]_{11}+[\cO_{\phi\ell}^{(3)}]_{22}-3[\cO_{\phi\ell}^{(1)}]_{22}+[\cO_{\phi\ell}^{(3)}]_{33}-3[\cO_{\phi\ell}^{(1)}]_{33}\Big]$
\\
\midrule
\midrule
\multirow{2}{*}{\vspace{-2.2cm}$\cB\sim(\bm1,\bm1)_0$}
&\multirow{3}{*}{\vspace{-0.5cm}$\bm3_1$}
&$\frac{(g_\cB^\ell)^2}{9M^2_\cB}\Big[-8[\cO_{\ell\ell}]_{1111}-2[\cO_{\ell\ell}]_{2222}-2[\cO_{\ell\ell}]_{3333}+8[\cO_{\ell\ell}]_{1122}+8[\cO_{\ell\ell}]_{1133}-4[\cO_{\ell\ell}]_{2233}$
\\
&&$-2[\cO_{\ell\ell}]_{1221}+2[\cO_{\ell\ell}]_{1331}-8[\cO_{\ell\ell}]_{2332}\Big]+\frac{(g_\cB^\ell)^2}{9M^2_\cB}\Big[-[\cO_{\ell\ell}]_{1212}-[\cO_{\ell\ell}]_{1313}-4[\cO_{\ell\ell}]_{2323}$
\\
&&$-2[\cO_{\ell\ell}]_{1213}+4[\cO_{\ell\ell}]_{1223}-2[\cO_{\ell\ell}]_{1231}+4[\cO_{\ell\ell}]_{1232}+4[\cO_{\ell\ell}]_{1323}+4[\cO_{\ell\ell}]_{1332}+\hermc\Big]$
\\
\cmidrule{2-3}
&\multirow{2}{*}{\vspace{-0.3cm}$\bm3_2$}
&\vspace{+0.1cm}$\frac{(g_\cB^\ell)^2}{2M^2_\cB}\Big[-[\cO_{\ell\ell}]_{2222}-[\cO_{\ell\ell}]_{3333}+2[\cO_{\ell\ell}]_{2233}-[\cO_{\ell\ell}]_{1331}-[\cO_{\ell\ell}]_{1221}\Big]$
\\
&&$+\frac{(g_\cB^\ell)^2}{4M^2_\cB}\Big[2[\cO_{\ell\ell}]_{1213}+2[\cO_{\ell\ell}]_{1231}-[\cO_{\ell\ell}]_{1212}-[\cO_{\ell\ell}]_{1313}+\hermc\Big]~~~~~~~~~~~$
\\
\midrule
\multirow{8}{*}{\vspace{-7.5cm}$\cW\sim(\bm1,\bm3)_0$}
&\multirow{2}{*}{\vspace{-0.37cm}$\bm1$}
&\vspace{+0.15cm}$-\frac{(g_\cW^\ell)^2}{8M^2_\cW}\Big[[\cO_{\ell\ell}]_{1111}+[\cO_{\ell\ell}]_{2222}+[\cO_{\ell\ell}]_{3333}+2[\cO_{\ell\ell}]_{1122}+2[\cO_{\ell\ell}]_{1133}+2[\cO_{\ell\ell}]_{2233}\Big]$
\\
&&$+\frac{(g_\cW^\ell)^2}{4M^2_\cW}\Big[-2[\cO_{\ell\ell}]_{1221}-2[\cO_{\ell\ell}]_{1331}-2[\cO_{\ell\ell}]_{2332}\Big]~~~~~~~~~~~~~~~~~~~~~~~~~~~~~~~~~~~~~~~~$
\\
\cmidrule{2-3}
&\multirow{3}{*}{\vspace{-0.8cm}$~\bm1'$}
&\vspace{+0.15cm}$\frac{(g_\cW^\ell)^2}{8M^2_\cW}\Big[2[\cO_{\ell\ell}]_{1221}+2[\cO_{\ell\ell}]_{1331}+[\cO_{\ell\ell}]_{2332}-4[\cO_{\ell\ell}]_{1122}-4[\cO_{\ell\ell}]_{2233}-4[\cO_{\ell\ell}]_{1133}\Big]$
\\
&&\vspace{+0.15cm}$+\frac{(g_\cW^\ell)^2}{8M^2_\cW}\Big[-4[\cO_{\ell\ell}]_{1123}-[\cO_{\ell\ell}]_{1212}-2[\cO_{\ell\ell}]_{1213}+2[\cO_{\ell\ell}]_{1223}+2[\cO_{\ell\ell}]_{1231}-2[\cO_{\ell\ell}]_{1232}$
\\
&&$-4[\cO_{\ell\ell}]_{1233}-[\cO_{\ell\ell}]_{1313}-4[\cO_{\ell\ell}]_{1322}-2[\cO_{\ell\ell}]_{1323}+2[\cO_{\ell\ell}]_{1332}-[\cO_{\ell\ell}]_{2323}+\hermc\Big]~~~~~$
\\
\cmidrule{2-3}
&\multirow{3}{*}{\vspace{-0.8cm}$~\,\bm1''$}
&\vspace{+0.15cm}$\frac{(g_\cW^\ell)^2}{8M^2_\cW}\Big[2[\cO_{\ell\ell}]_{1221}+2[\cO_{\ell\ell}]_{1331}+[\cO_{\ell\ell}]_{2332}-4[\cO_{\ell\ell}]_{1122}-4[\cO_{\ell\ell}]_{2233}-4[\cO_{\ell\ell}]_{1133}\Big]$
\\
&&\vspace{+0.15cm}$+\frac{(g_\cW^\ell)^2}{8M^2_\cW}\Big[-4[\cO_{\ell\ell}]_{1123}-[\cO_{\ell\ell}]_{1212}-2[\cO_{\ell\ell}]_{1213}+2[\cO_{\ell\ell}]_{1223}+2[\cO_{\ell\ell}]_{1231}-2[\cO_{\ell\ell}]_{1232}$
\\
&&$-4[\cO_{\ell\ell}]_{1233}-[\cO_{\ell\ell}]_{1313}-4[\cO_{\ell\ell}]_{1322}-2[\cO_{\ell\ell}]_{1323}+2[\cO_{\ell\ell}]_{1332}-[\cO_{\ell\ell}]_{2323}+\hermc\Big]~~~~~$
\\
\cmidrule{2-3}
&\multirow{4}{*}{\vspace{-0.85cm}$~\bm3_1$}
&$~~~~~~\frac{(g_\cW^\ell)^2}{36M^2_\cW}\Big[-8[\cO_{\ell\ell}]_{1111}-2[\cO_{\ell\ell}]_{2222}-2[\cO_{\ell\ell}]_{3333}-12[\cO_{\ell\ell}]_{1122}-12[\cO_{\ell\ell}]_{1133}-12[\cO_{\ell\ell}]_{2233}$
\\
&&\vspace{+0.15cm}$~~~~~~+18[\cO_{\ell\ell}]_{1221}+18[\cO_{\ell\ell}]_{1331}\Big]+\frac{(g_\cW^\ell)^2}{36M^2_\cW}\Big[-4[\cO_{\ell\ell}]_{1123}-[\cO_{\ell\ell}]_{1212}-2[\cO_{\ell\ell}]_{1213}-4[\cO_{\ell\ell}]_{1223}$
\\
&&\vspace{+0.15cm}$~~~~~+2[\cO_{\ell\ell}]_{1231}+4[\cO_{\ell\ell}]_{1232}+4[\cO_{\ell\ell}]_{1233}-[\cO_{\ell\ell}]_{1313}+8[\cO_{\ell\ell}]_{1322}+4[\cO_{\ell\ell}]_{1323}-4[\cO_{\ell\ell}]_{1332}$
\\&&$-4[\cO_{\ell\ell}]_{2323}+\hermc\Big]~~~~~~~~~~~~~~~~~~~~~~~~~~~~~~~~~~~~~~~~~~~~~~~~~~~~~~~~~~~~~~~~~~~~~~~~~~~~~~~~~$
\\
\cmidrule{2-3}
&\multirow{3}{*}{\vspace{-0.35cm}$~\bm3_2$}
&$\frac{(g_\cW^\ell)^2}{16M^2_\cW}\Big[2[\cO_{\ell\ell}]_{1221}+2[\cO_{\ell\ell}]_{1331}-4[\cO_{\ell\ell}]_{1122}-4[\cO_{\ell\ell}]_{1133}-4[\cO_{\ell\ell}]_{2233}-2[\cO_{\ell\ell}]_{3333}$
\\
&&$-2[\cO_{\ell\ell}]_{2222}+8[\cO_{\ell\ell}]_{2332}\Big]+\frac{(g_\cW^\ell)^2}{16M^2_\cW}\Big[4[\cO_{\ell\ell}]_{1123}-[\cO_{\ell\ell}]_{1212}+2[\cO_{\ell\ell}]_{1213}-2[\cO_{\ell\ell}]_{1231}$
\\
&&$-[\cO_{\ell\ell}]_{1313}+\hermc\Big]~~~~~~~~~~~~~~~~~~~~~~~~~~~~~~~~~~~~~~~~~~~~~~~~~~~~~~~~~~~~~~~~~~~~~~~~~~~~~~~~~~$
\\
\bottomrule
\end{tabular}
}
\caption{Tree-level matching onto dimension-6 SMEFT operators for NP mediators generating \textit{discrete leptonic directions} under $A_4$ flavor symmetry assumptions. The SMEFT operators appearing in the matching relations are defined in Tab.~\ref{tab:SMEFToperators}.}
\label{tab:A4 SMEFT Matching}
\end{table*}

\begin{table*}[t]
\centering
\scalebox{1}{
\begin{tabular}{ccc}
\toprule
\multicolumn{3}{c}{\textbf{$A_5$ SMEFT Matching}}
\\
\midrule
\textbf{UV Field} & \textbf{Irrep}&\textbf{$\cL_{\sscript{SMEFT}}\supset$}
\\
\midrule
$\cS_1\sim(\bm1,\bm1)_{1}$
&$\bm3$
&$\frac{|y_{\cS_1}|^2}{2M^2_{\cS_1}}\Big[ [\cO_{\ell\ell}]_{1122}-[\cO_{\ell\ell}]_{1221}+[\cO_{\ell\ell}]_{1133}
			-[\cO_{\ell\ell}]_{1331}+[\cO_{\ell\ell}]_{2233}-[\cO_{\ell\ell}]_{2332} \Big]$
\\
\midrule
\multirow{2}{*}{\vspace{-1.8cm}$\Xi_1\sim(\bm1,\bm3)_{1}$}
&$\bm1$
&\vspace{+0.1cm}$\frac{|y_{\Xi_1}|^2}{M^2_{\Xi_1}}\Big[ [\cO_{\ell\ell}]_{1111}+2[\cO_{\ell\ell}]_{2233}+2[\cO_{\ell\ell}]_{2332} +(2[\cO_{\ell\ell}]_{1213}+\hermc)\Big]$
\\
\cmidrule{2-3}
&\multirow{2}{*}{\vspace{-0.5cm}$\bm5$}
&\vspace{+0.1cm}$\frac{|y_{\Xi_1}|^2}{M^2_{\Xi_1}}\Big[4[\cO_{\ell\ell}]_{1111}+6[\cO_{\ell\ell}]_{2222}+6[\cO_{\ell\ell}]_{3333}+6[\cO_{\ell\ell}]_{1122}+6[\cO_{\ell\ell}]_{1133}~~~~~~$
\\
&
&\vspace{+0.1cm}$~~~~~~~~~+2[\cO_{\ell\ell}]_{2233}+6[\cO_{\ell\ell}]_{1221}+6[\cO_{\ell\ell}]_{1331}+2[\cO_{\ell\ell}]_{2332}\Big]-\frac{|y_{\Xi_1}|^2}{M^2_{\Xi_1}}\Big[4[\cO_{\ell\ell}]_{1213}+\hermc\Big]$
\\
\midrule
\midrule
$N\sim(\bm1,\bm1)_0$
&$\bm3$
&\vspace{+0.1cm}$\frac{|\lambda_N|^2}{4M^2_N}\Big[[\cO_{\phi\ell}^{(1)}]_{11}-[\cO_{\phi\ell}^{(3)}]_{11}+[\cO_{\phi\ell}^{(1)}]_{22}-[\cO_{\phi\ell}^{(3)}]_{22}+[\cO_{\phi\ell}^{(1)}]_{33}-[\cO_{\phi\ell}^{(3)}]_{33}\Big]$
\\
\midrule
$E\sim(\bm1,\bm1)_{-1}$
&$\bm3$
&$-\frac{|\lambda_E|^2}{4M^2_E}\Big[[\cO_{\phi\ell}^{(1)}]_{11}+[\cO_{\phi\ell}^{(3)}]_{11}+[\cO_{\phi\ell}^{(1)}]_{22}+[\cO_{\phi\ell}^{(3)}]_{22}+[\cO_{\phi\ell}^{(1)}]_{33}+[\cO_{\phi\ell}^{(3)}]_{33}\Big]$
\\
\midrule
$\Sigma\sim(\bm1,\bm3)_{0}$
&$\bm3$
&$\frac{|\lambda_\Sigma|^2}{16M^2_\Sigma}\Big[[\cO_{\phi\ell}^{(3)}]_{11}+3[\cO_{\phi\ell}^{(1)}]_{11}+[\cO_{\phi\ell}^{(3)}]_{22}+3[\cO_{\phi\ell}^{(1)}]_{22}+[\cO_{\phi\ell}^{(3)}]_{33}+3[\cO_{\phi\ell}^{(1)}]_{33}\Big]$
\\
\midrule
$\Sigma_1\sim(\bm1,\bm3)_{-1}$
&$\bm3$
&$\frac{|\lambda_{\Sigma_1}|^2}{16M^2_{\Sigma_1}}\Big[[\cO_{\phi\ell}^{(3)}]_{11}-3[\cO_{\phi\ell}^{(1)}]_{11}+[\cO_{\phi\ell}^{(3)}]_{22}-3[\cO_{\phi\ell}^{(1)}]_{22}+[\cO_{\phi\ell}^{(3)}]_{33}-3[\cO_{\phi\ell}^{(1)}]_{33}\Big]$
\\
\midrule
\midrule
\multirow{2}{*}{\vspace{-1.8cm}$\cB\sim(\bm1,\bm1)_0$}
&\multirow{2}{*}{\vspace{-0.5cm}$\bm3$}
&\vspace{+0.1cm}$-\frac{2(g_\cB^\ell)^2}{M^2_\cB}\Big[[\cO_{\ell\ell}]_{1221}+[\cO_{\ell\ell}]_{1331}+[\cO_{\ell\ell}]_{2222}+[\cO_{\ell\ell}]_{3333}+2[\cO_{\ell\ell}]_{2233}\Big]$
\\
&&$+\frac{(g_\cB^\ell)^2}{M^2_\cB}\Big[2[\cO_{\ell\ell}]_{1213}+2[\cO_{\ell\ell}]_{1231}-[\cO_{\ell\ell}]_{1212}-[\cO_{\ell\ell}]_{1313}+\hermc\Big]~~~~$
\\
\cmidrule{2-3}
&\multirow{2}{*}{\vspace{-0.3cm}$\bm5$}
&\vspace{+0.1cm}$-\frac{4(g_\cB^\ell)^2}{M^2_\cB}\Big[2[\cO_{\ell\ell}]_{1111}+3[\cO_{\ell\ell}]_{1331}+[\cO_{\ell\ell}]_{2332}+3[\cO_{\ell\ell}]_{3333}\Big]$
\\
&&$+\frac{(g_\cB^\ell)^2}{M^2_\cB}\Big[8[\cO_{\ell\ell}]_{1123}-6[\cO_{\ell\ell}]_{1313}-2[\cO_{\ell\ell}]_{2323}+\hermc\Big]~~~~~$
\\
\midrule
\multirow{7}{*}{\vspace{-2.4cm}$\cW\sim(\bm1,\bm3)_0$}
&\multirow{2}{*}{\vspace{-0.37cm}$\bm1$}
&\vspace{+0.15cm}$-\frac{(g_\cW^\ell)^2}{8M^2_\cW}\Big[[\cO_{\ell\ell}]_{1111}+[\cO_{\ell\ell}]_{2222}+[\cO_{\ell\ell}]_{3333}+2[\cO_{\ell\ell}]_{1122}+2[\cO_{\ell\ell}]_{1133}+2[\cO_{\ell\ell}]_{2233}\Big]$
\\
&&$+\frac{(g_\cW^\ell)^2}{4M^2_\cW}\Big[-2[\cO_{\ell\ell}]_{1221}-2[\cO_{\ell\ell}]_{1331}-2[\cO_{\ell\ell}]_{2332}\Big]~~~~~~~~~~~~~~~~~~~~~~~~~~~~~~~~~~~~~~~~$
\\
\cmidrule{2-3}
&\multirow{4}{*}{\vspace{-0.4cm}$\bm3$}
&\vspace{+0.15cm}$-\frac{(g_\cW^\ell)^2}{2M^2_\cW}\Big[2[\cO_{\ell\ell}]_{1122}+2[\cO_{\ell\ell}]_{1133}+2[\cO_{\ell\ell}]_{2233}+[\cO_{\ell\ell}]_{2222}+[\cO_{\ell\ell}]_{3333}\Big]~~~~~~$
\\
&&\vspace{+0.15cm}$~~~~~+\frac{(g_\cW^\ell)^2}{4M^2_\cW}\Big[2[\cO_{\ell\ell}]_{1221}+2[\cO_{\ell\ell}]_{1331}+8[\cO_{\ell\ell}]_{2332}\Big]+\frac{(g_\cW^\ell)^2}{4M^2_\cW}\Big[-[\cO_{\ell\ell}]_{1313}-[\cO_{\ell\ell}]_{1212}$
\\
&&\vspace{+0.15cm}$+4[\cO_{\ell\ell}]_{1123}+2[\cO_{\ell\ell}]_{1213}-2[\cO_{\ell\ell}]_{1231}+\hermc\Big]~~~~~~~~~~~~~~~~~~~~~~~~~~~~~~~~~~~~$
\\
\cmidrule{2-3}
&\multirow{3}{*}{\vspace{-0.35cm}$\bm5$}
&\vspace{+0.15cm}$~~~~~~~~\frac{(g_\cW^\ell)^2}{M^2_\cW}\Big[-2[\cO_{\ell\ell}]_{1111}-6[\cO_{\ell\ell}]_{1133}+3[\cO_{\ell\ell}]_{1331}-3[\cO_{\ell\ell}]_{3333}+[\cO_{\ell\ell}]_{2332}-2[\cO_{\ell\ell}]_{2233}\Big]$
\\
&&\vspace{+0.15cm}$+\frac{(g_\cW^\ell)^2}{2M^2_\cW}\Big[-4[\cO_{\ell\ell}]_{1123}+8[\cO_{\ell\ell}]_{1231}-3[\cO_{\ell\ell}]_{1313}-[\cO_{\ell\ell}]_{2323}+\hermc\Big]~~~~~~~~~~~~~~~~~~$
\\
\bottomrule
\end{tabular}
}
\caption{Tree-level matching onto dimension-6 SMEFT operators for NP mediators generating \textit{discrete leptonic directions} under $A_5$ flavor symmetry assumptions. The SMEFT operators appearing in the matching relations are defined in Tab.~\ref{tab:SMEFToperators}.}
\label{tab:A5 SMEFT Matching}
\end{table*}

\begin{table*}[t]
\centering
\scalebox{1.00}{
\begin{tabular}{ccc}
\toprule
\multicolumn{3}{c}{\textbf{$S_4$ SMEFT Matching}}
\\
\midrule
\textbf{UV Field} & \textbf{Irrep}&\textbf{$\cL_{\sscript{SMEFT}}\supset$}
\\
\midrule
$\cS_1\sim(\bm1,\bm1)_{1}$
&$\bm3$
&$\frac{|y_{\cS_1}|^2}{2M^2_{\cS_1}}\Big[ [\cO_{\ell\ell}]_{1122}-[\cO_{\ell\ell}]_{1221}+[\cO_{\ell\ell}]_{1133}
			-[\cO_{\ell\ell}]_{1331}+[\cO_{\ell\ell}]_{2233}-[\cO_{\ell\ell}]_{2332} \Big]$
\\
\midrule
\multirow{2}{*}{\vspace{-1.7cm}$\Xi_1\sim(\bm1,\bm3)_{1}$}
&$\bm1$
&\vspace{+0.1cm}$\frac{|y_{\Xi_1}|^2}{M^2_{\Xi_1}}\Big[[\cO_{\ell\ell}]_{1111}+[\cO_{\ell\ell}]_{2222}+[\cO_{\ell\ell}]_{3333}+([\cO_{\ell\ell}]_{1212}+[\cO_{\ell\ell}]_{1313}+[\cO_{\ell\ell}]_{2323}+\hermc)\Big]$
\\
\cmidrule{2-3}
&\multirow{1}{*}{\vspace{-0.0cm}$\bm2$}
&\vspace{+0.0cm}$\frac{2|y_{\Xi_1}|^2}{3M^2_{\Xi_1}}\Big[[\cO_{\ell\ell}]_{1111}+[\cO_{\ell\ell}]_{2222}+[\cO_{\ell\ell}]_{3333}\Big]-\frac{|y_{\Xi_1}|^2}{3M^2_{\Xi_1}}\Big[ [\cO_{\ell\ell}]_{1212}+[\cO_{\ell\ell}]_{1313}+[\cO_{\ell\ell}]_{2323}+\hermc \Big]$
\\
\cmidrule{2-3}
&$\bm3$
&$\frac{2|y_{\Xi_1}|^2}{M^2_{\Xi_1}}\Big[ [\cO_{\ell\ell}]_{1122}+[\cO_{\ell\ell}]_{1133}+[\cO_{\ell\ell}]_{2233}+[\cO_{\ell\ell}]_{1221}+[\cO_{\ell\ell}]_{1331}+[\cO_{\ell\ell}]_{2332} \Big]$
\\
\midrule
\midrule
$N\sim(\bm1,\bm1)_0$
&$\bm3$
&\vspace{+0.1cm}$\frac{|\lambda_N|^2}{4M^2_N}\Big[[\cO_{\phi\ell}^{(1)}]_{11}-[\cO_{\phi\ell}^{(3)}]_{11}+[\cO_{\phi\ell}^{(1)}]_{22}-[\cO_{\phi\ell}^{(3)}]_{22}+[\cO_{\phi\ell}^{(1)}]_{33}-[\cO_{\phi\ell}^{(3)}]_{33}\Big]$
\\
\midrule
$E\sim(\bm1,\bm1)_{-1}$
&$\bm3$
&$-\frac{|\lambda_E|^2}{4M^2_E}\Big[[\cO_{\phi\ell}^{(1)}]_{11}+[\cO_{\phi\ell}^{(3)}]_{11}+[\cO_{\phi\ell}^{(1)}]_{22}+[\cO_{\phi\ell}^{(3)}]_{22}+[\cO_{\phi\ell}^{(1)}]_{33}+[\cO_{\phi\ell}^{(3)}]_{33}\Big]$
\\
\midrule
$\Sigma\sim(\bm1,\bm3)_{0}$
&$\bm3$
&$\frac{|\lambda_\Sigma|^2}{16M^2_\Sigma}\Big[[\cO_{\phi\ell}^{(3)}]_{11}+3[\cO_{\phi\ell}^{(1)}]_{11}+[\cO_{\phi\ell}^{(3)}]_{22}+3[\cO_{\phi\ell}^{(1)}]_{22}+[\cO_{\phi\ell}^{(3)}]_{33}+3[\cO_{\phi\ell}^{(1)}]_{33}\Big]$
\\
\midrule
$\Sigma_1\sim(\bm1,\bm3)_{-1}$
&$\bm3$
&$\frac{|\lambda_{\Sigma_1}|^2}{16M^2_{\Sigma_1}}\Big[[\cO_{\phi\ell}^{(3)}]_{11}-3[\cO_{\phi\ell}^{(1)}]_{11}+[\cO_{\phi\ell}^{(3)}]_{22}-3[\cO_{\phi\ell}^{(1)}]_{22}+[\cO_{\phi\ell}^{(3)}]_{33}-3[\cO_{\phi\ell}^{(1)}]_{33}\Big]$
\\
\midrule
\midrule
\multirow{3}{*}{\vspace{-0.1cm}$\cB\sim(\bm1,\bm1)_0$}
&\multirow{1}{*}{\vspace{-0.0cm}$\bm2$}
&\vspace{+0.1cm}$-\frac{(g_\cB^\ell)^2}{3M^2_\cB}\Big[[\cO_{\ell\ell}]_{1111}+[\cO_{\ell\ell}]_{2222}+[\cO_{\ell\ell}]_{3333}-[\cO_{\ell\ell}]_{1122}-[\cO_{\ell\ell}]_{1133}-[\cO_{\ell\ell}]_{2233}\Big]$
\\
\cmidrule{2-3}
&\multirow{1}{*}{\vspace{-0.0cm}$\bm3$}
&\vspace{+0.1cm}$-\frac{(g_\cB^\ell)^2}{2M^2_\cB}\Big[ 2[\cO_{\ell\ell}]_{1221}+2[\cO_{\ell\ell}]_{1331}+2[\cO_{\ell\ell}]_{2332}+([\cO_{\ell\ell}]_{1212}+[\cO_{\ell\ell}]_{1313}+[\cO_{\ell\ell}]_{2323}+\hermc) \Big]$
\\
\midrule
\multirow{7}{*}{\vspace{-1.7cm}$\cW\sim(\bm1,\bm3)_0$}
&\multirow{2}{*}{\vspace{-0.37cm}$\bm1$}
&\vspace{+0.15cm}$-\frac{(g_\cW^\ell)^2}{8M^2_\cW}\Big[[\cO_{\ell\ell}]_{1111}+[\cO_{\ell\ell}]_{2222}+[\cO_{\ell\ell}]_{3333}-2[\cO_{\ell\ell}]_{1122}-2[\cO_{\ell\ell}]_{1133}-2[\cO_{\ell\ell}]_{2233}$
\\
&&$+4[\cO_{\ell\ell}]_{1221}+4[\cO_{\ell\ell}]_{1331}+4[\cO_{\ell\ell}]_{2332}\Big]~~~~~~~~~~~~~~~~~~~~~~~~~~~~~~~~~~~~~~~~~~~~~~~~~~~$
\\
\cmidrule{2-3}
&\multirow{2}{*}{\vspace{-0.4cm}$\bm2$}
&\vspace{+0.15cm}$-\frac{(g_\cW^\ell)^2}{12M^2_\cW}\Big[[\cO_{\ell\ell}]_{1111}+[\cO_{\ell\ell}]_{2222}+[\cO_{\ell\ell}]_{3333}+[\cO_{\ell\ell}]_{1122}+[\cO_{\ell\ell}]_{1133}+[\cO_{\ell\ell}]_{2233}$
\\
&&\vspace{+0.15cm}$-2[\cO_{\ell\ell}]_{1221}-2[\cO_{\ell\ell}]_{1331}-2[\cO_{\ell\ell}]_{2332}\Big]~~~~~~~~~~~~~~~~~~~~~~~~~~~~~~~~~~~~~~~~~~~~~~~$
\\
\cmidrule{2-3}
&\multirow{3}{*}{\vspace{-0.35cm}$\bm3$}
&\vspace{+0.15cm}$-\frac{(g_\cW^\ell)^2}{4M^2_\cW}\Big[2[\cO_{\ell\ell}]_{1122}+2[\cO_{\ell\ell}]_{1133}+2[\cO_{\ell\ell}]_{2233}-[\cO_{\ell\ell}]_{1221}-[\cO_{\ell\ell}]_{2332}-[\cO_{\ell\ell}]_{1331}\Big]$
\\
&&\vspace{+0.15cm}$-\frac{(g_\cW^\ell)^2}{8M^2_\cW}\Big[[\cO_{\ell\ell}]_{1212}+[\cO_{\ell\ell}]_{1313}+[\cO_{\ell\ell}]_{2323}+\hermc\Big]~~~~~~~~~~~~~~~~~~~~~~~~~~~~~~~~~~~~~~~~$
\\
\bottomrule
\end{tabular}
}
\caption{Tree-level matching onto dimension-6 SMEFT operators for NP mediators generating \textit{discrete leptonic directions} under $S_4$ flavor symmetry assumptions. The SMEFT operators appearing in the matching relations are defined in Tab.~\ref{tab:SMEFToperators}.}
\label{tab:S4 SMEFT Matching}
\end{table*}

\clearpage
\bibliography{letter}

\end{document}